\documentclass[twocolumn,floatfix,prl,aps,showpacs]{revtex4-2}
\usepackage{graphicx,amsmath,amssymb,color}
\usepackage[utf8]{inputenc}
\usepackage{nicefrac}
\usepackage{multirow, makecell, ragged2e}
\usepackage[titletoc,title]{appendix}
\usepackage{hyperref}

\newcommand{\be}{\begin{equation}}
\newcommand{\ee}{\end{equation}}

\newcommand{\ba}{\begin{eqnarray}}
\newcommand{\ea}{\end{eqnarray}}

\begin{document}

\title{Characterizing the many-body localization crossover as a metal-insulator transition: localization length from polarization and quantum metric}
\author{W. N. Faugno, Tomoki Ozawa}
\affiliation{Advanced Institute for Materials Research (WPI-AIMR), Tohoku University, Sendai 980-8577, Japan}
\date{\today}

\begin{abstract}
Many-body localization (MBL) represents a unique physical phenomenon, providing a testing ground for exploring thermalization, or more precisely its failure. Here we characterize the MBL regime geometrically by the many-body quantum metric (MBQM), defined in the parameter space of twist boundary, and the localization parameter as defined in the modern theory of polarization and insulators. First, we demonstrate that the quantum metric can be used to characterize disordered insulating states by applying this theoretical framework to excited states of the one-dimensional (1D) Anderson insulator. There we observe that the MBQM and localization parameter are related in finite realizations despite the states being gapless in the thermodynamic limit. Then, we consider a disordered 1D Bose-Hubbard model and find that we can characterize the ergodic-MBL crossover by comparing the MBQM and localization parameter. We find that we can extract a natural localization length in the MBL regime that characterizes the real space spread of the wave function and can be measured by extracting the quantum metric. Our analysis provides complementary insight into the MBL regime focusing on its insulating properties and providing a localization length whose definition is consistent across a range of insulating phases.
\end{abstract}

\maketitle
\section{Introduction}
Advancements in highly controllable experimental platforms, such as optical lattices\cite{Bloch08,Bilokon23,Chomaz23,Malz23}, trapped ions\cite{Blatt12}, and Rydberg atom arrays\cite{Adams20,Wu21}, have allowed for realizations of well isolated quantum systems in the laboratory where quantum dynamics and nonequilibrium physics can be observed \cite{Kinoshita06,Vasseur16}. Of particular interest are nonergodic systems, which do not thermalize within a finite time. These systems can host stable phases of matter and help to understand thermalization mechanisms \cite{Nandkishore15}. Thus far the only potential robust nonergodic phase known is many-body localization (MBL), which arises from the interplay of disorder and interactions \cite{DeLuca13,Levi16}. In contrast, other nonergodic phases, such as noninteracting phases or integrable models, either require fine tuning \cite{DAlessio16} or only exhibit non thermalization for a small subset of configurations as in quantum scars~\cite{Turner18}.

Anderson localization was the first identified nonergodic system wherein disorder in a noninteracting system localizes all eigenstates \cite{Anderson58}. Following studies saw the inclusion of interactions to determine the stability of the ground state in this localized phase, finding that indeed localization persisted in several regimes \cite{Fleishman80,Finkelstein83,Giamarchi88}. More recent studies have found that excited states, even those corresponding to infinite temperature, can remain localized, resulting in a phenomenon termed 
MBL~\cite{Altshuler97,Gornyi05,Basko06}.

MBL hosts many interesting properties beyond those observed in its noninteracting counterpart and demonstrates a unique dynamical phase transition between the localized and delocalized regions. The phase boundary can be determined through how the entanglement entropy of excited states scales with system size, growing with volume in the delocalized phase and following an area law in the localized phase \cite{Bauer13}. Additionally the entanglement entropy of initial product states has been shown to grow logarithmically, in contrast to the quickly saturating entanglement entropy in noninteracting Anderson insulators \cite{Serbyn13,Huang21}. The unique behavior of the entanglement entropy is related to the localization in Fock space in addition to real space localization. This has been shown to be related to the emergence of a set of quasilocal integrals of motion (LIOM), which has allowed for the construction of phenomenological models of MBL\cite{Abanin19}. Strong evidence of MBL has been found in the disordered Bose-Hubbard \cite{Sierant18,Lukin19} and Fermi-Hubbard models \cite{BarLev15} in one-dimensional (1D) as well as quasidisordered lattice systems with non commensurate periodic external potentials \cite{Schreiber15}. 

The existence of MBL as a thermodynamic phase has been called into question as it has been proposed that generically a region of low disorder can appear in the thermodynamic limit and act as a thermal bath, promoting thermalization of the entire system \cite{Sels21,Kiefer21,Sels23}. It remains an open question if MBL truly represents a thermodynamic phase or is restricted to being a dynamical phase, only observable in finite systems. Further confounding the issue, numerical studies have been unable to convincingly determine the critical disorder at which the MBL transition occurs~\cite{Sierant25}. System size dependent drift of the critical disorder has been observed in such key indicators as the level spacing statistics and entanglement entropy~\cite{Sierant20}. While the fate of MBL in the thermodynamic limit remains an open question, it still appears to survive as a dynamical phase and can be observed in finite size systems in experiments~\cite{Schreiber15,Luschen17,Lukin19,Rispoli19,Leonard23}.

An important parameter in understanding insulating states is the characteristic localization length. To present, there have been several proposed approaches to defining the localization length in MBL systems. For example, a localization length can be defined in terms of the inverse participation ratio obtained from the single particle density matrix \cite{Bera15}. Another proposal has shown that a localization length can be extracted from the decay of the spatial correlation function~\cite{Pal10}. Localization has also been defined by examining the operator growth~\cite{Weisse25}. Other methods rely on phenomenological models constructed from the extensive set of LIOM, which introduces a hierarchy of length scales\cite{Serbyn13b,Huse14}. The LIOM have been constructed explicitly for the XXZ model \cite{Chandran15}, the disordered hardcore Fermi-Hubbard model \cite{Rademaker16,Thomson18}, and the Heisenberg model\cite{Pekker17}. Although there are several different localization lengths that appear in MBL, their physical meaning or experimental signature are often unclear.
 
A natural localization length has been defined in the modern theory of insulators, built upon the theory of polarization\cite{Resta98,Resta99,Souza00,Resta02,Lu10,Valenca19,Hetenyi19,Hetenyi22}. This localization length allows us to distinguish conducting and insulating phases as it diverges in the former, but remains finite in the latter. Additionally, for insulators, this localization length can be related to the so-called quantum metric, which defines the distance between wave functions in the space of some parameter of the Hamiltonian. Typical choices of such parameters are the crystal momentum or the twist boundary phase. he localization length is related to the many-body quantum metric (MBQM) defined in the space of twist boundary conditions. This formalism has been applied in several paradigmatic contexts, including quantum Hall insulators\cite{Resta05}, Mott insulators\cite{Wilkens01}, and Chern insulators\cite{Thonhauser06} and even in more exotic contexts like quasicrystals~\cite{Marsal25}. Twist boundary condition was used to characterize the MBL regime through the Drude weight~\cite{Filippone16}, which also captures the metal-insulator transition. The boundary condition sensitivity of energy levels has shown to characterize both Anderson insulators and MBL regimes~\cite{Pouranvari21}. The MBQM is defined in terms of the change in the wave function and energy in response to a change in the twist boundary condition or equivalently the response to a change in the vector potential. In MBL, a similar concept has been used to define the localization length in terms of response to an imaginary vector potential~\cite{Hamazaki19,Heuben21,Obrien23}. From a more general perspective, the MBQM is a particular form of the fidelity susceptibility, which has been used to characterize the MBL through level spacings from their connection to random matrix theory~\cite{Sierant19,Maksymov19,Sels21}.

In this work, we apply the formalism provided by the modern theory of insulators to highly excited states of the disordered hardcore Bose-Hubbard model. While this formalism usually requires that the state be gapped, our results demonstrate that even though these states are gapless in the thermodynamic limit, we can distinguish the ergodic and MBL regimes by using the fact that the localization parameter and quantum metric are unrelated for conducting states, but are approximately equal in insulators for finite systems, converging to the same value in the thermodynamic limit. In finite systems, the quantum metric can still be defined as there is generically an energy difference between states for any particular disorder configuration. We introduce a parameter to measure the agreement between these two quantities, such that this parameter goes to one deep in the ergodic regime and zero in the MBL regime. We show that this parameter reasonably estimates the ergodic to MBL transition in finite size systems and shows the characteristic drift of the critical disorder with system size. Finally, we define a characteristic localization length in the MBL regime as prescribed by the modern theory of insulators. Our results thus provide another parameter by which MBL can be characterized as well as a natural localization length.

\section{The Modern theory of polarization and the quantum metric}
We review the definitions of the localization length in periodic boundary conditions (PBC) as originally proposed by Resta~\cite{Resta98} and the MBQM, the two key quantities in this work. The localization length of a many-body state $\Psi$ under PBC is built on the quantity, $D_N$, which we call the localization parameter,
\begin{align}
    D_N =& -N \ln |z_N|^2\\
    z_N =& \langle \Psi | e^{2i\pi X_{CM}/L} | \Psi \rangle
\end{align}
where $N$ is the number of particles, $X_{CM} \equiv \sum_n^N x_n$ with $x_n$ the position operator for the $n$-th particle, and $L$ is the length of the system. The quantity $z_N$ defines both localization and polarization properties of systems under PBC where the position operator is ill-defined. Transport properties are captured by the thermodynamic limit $D_\infty = \lim_{N\rightarrow\infty} D_N$, which has been shown to discriminate between insulators and conductors: it remains finite in the former, but diverges in the latter with increasing system size\cite{Resta98,Resta11}. This property is known to hold for both band and disorder induced insulators and for both interacting and noninteracting systems. For an insulator in 1D, we define the localization length $\ell_N = \sqrt{D_N}/2\pi n$ where $n$ is the density, $n=N/L$\cite{Resta20}. Thus, this quantity provides a general theoretical framework for probing conduction properties and defining a natural localization length.

For an insulator, the localization parameter, $D_N$, defined above is approximately proportional to the MBQM in finite systems and converges to the same value in the thermodynamic limit. The MBQM is a geometrical property of many-body quantum states that describes the change in the state with regards to some parameter of the Hamiltonian. Here we focus on the MBQM in response to changes in the twist boundary condition. We consider 1D systems in this article where the MBQM is equivalent to the many-body quantum geometric tensor. Twisted boundary conditions with phase $\theta$ are equivalent to inserting a flux $\theta$ through the system with the wavefunction obeying the standard PBC without twist~\cite{Niu85}. Let us denote the Hamiltonian of the system with a flux $\theta$ inserted by $H(\theta)$.
The MBQM, which we denote by $g(\phi_0,\theta)$, for a state $|\phi_0\rangle$, can then be defined in two equivalent ways~\cite{Banerjee96,Venuti07}
\begin{align}
    g(\phi_0,\theta) =& \langle\partial_\theta \phi_0(\theta)|\partial_\theta \phi_0(\theta)\rangle - |\langle \phi_0(\theta)|\partial_\theta \phi_0(\theta)\rangle|^2\label{eq:gmunu1}\\
    =& \sum_{n \neq 0} \frac{\langle \phi_0|\partial_{\theta} H (\theta) |\phi_n\rangle \langle \phi_n|\partial_{\theta} H (\theta) |\phi_0 \rangle}{(E_0 -E_n)^2},
    \label{eq:gmunu2}
\end{align}
where the sum is over all the eigenstates $|\phi_n \rangle$, whose corresponding eigenenergies are $E_n$, different from the state under consideration, $|\phi_0 \rangle$. (In general, both $|\phi_n \rangle$ and $E_n$ depend on $\theta$.) In this work we employ Eq.~\ref{eq:gmunu2} to calculate the MBQM. This formula is derived from Eq.~\ref{eq:gmunu1} under the assumption of an energy gap. While the excited states we are considering are gapless in the thermodynamic limit, in disorder systems the probability of an exact degeneracy is 0, which allows us to apply Eq.~\ref{eq:gmunu2} as we shall see below. When the system is large enough, we can expect that $g(\phi_0,\theta)$ does not depend on $\theta$ following arguments similar to those in Refs.~\cite{Niu85,Souza00,Watanabe18}, though this is only valid for a gapped system. 
In this work we consider fixed $\theta=0$ corresponding to a system at zero flux and will thus suppress the arguments in $g(\phi_0,\theta)$ to simplify notation. This is an important consideration since the excited states considered below are expected to undergo some level crossings when $\theta$ is varied. It has been shown that for a system of length $L$ the quantum metric is related to the variance in the center-of mass-coordinate as $g = \mathrm{Var}(X_{CM})/L^2$~\cite{Ozawa19}. This can be shown numerically when the wavefunction is fully localized within a finite region, i.e. the coordinates can be shifted so that the wave function does not cross the periodic boundary, so that the variance can be calculated in a well-defined manner~\cite{Souza00}.

In insulators, the MBQM and the localization parameter are related as
\begin{equation}
    D_N \simeq g_N =  4\pi^2Ng
\end{equation}
with equality in the thermodynamic limit~\cite{Resta20}. We have defined $g_N$ as the MBQM with appropriate prefactors, such that it coincides with the localization parameter in the thermodynamic limit. The localization length of an insulator can then be defined either by the MBQM or the localization parameter as $\ell_N = \sqrt{g_N}/2\pi n = \sqrt{D_N}/2\pi n$~\cite{Resta99}. We stress that we are using Resta's definition of the localization length, which can be interpreted as the variance of the center-of-mass coordinate for many-body systems even in the presence of interactions. These results are rigorously defined for gapped systems and may be expected to break down in gapless systems as can be seen from the denominator in Eq.\ref{eq:gmunu2}. From our numerical results below, we find that the MBQM can still be related to the localization in finite size disordered systems as for any finite sized disorder realization there is a 0 probability chance that there is an exact degeneracy. We also note that the quantum metric can still be defined in the presence of degeneracy~\cite{Ding24}. We summarize the expected behaviors of the localization parameter and MBQM in Table~\ref{tab:Summary} for conductors and insulators in finite systems with PBC as well as in the thermodynamic limit. Note that the fate of the MBQM in the thermodynamic limit is not fully determined for a conducting system, but depends on the conductor in question. For example, in a free fermion band conductor, the MBQM is expected to be finite~\cite{Ozawa21}.

\begin{table}
\centering
    \begin{tabular}{|c|c|c|}
    \hline
         & Conductor & Insulator\\
         \hline
        Finite Size System  & $D_N\neq g_N$ & $D_N\simeq g_N$ \\
        Thermodynamic Limit & $D_N\rightarrow\infty,g_N\rightarrow?$ & $D_N=g_N <\infty$ \\
        \hline
\end{tabular}
\caption{Summary of behaviors of the MBQM and localization parameter for different systems. Conductors and insulators can be distinguished by the agreement between the localization parameter and the MBQM or by the divergence of the localization parameter in the thermodynamic limit. In conductors, the fate of the MBQM in the thermodynamic limit depends on the particular system considered.}
\label{tab:Summary}
\end{table}

We apply this method to the disordered hardcore Bose-Hubbard chain to demonstrate that it captures the transition from the ergodic regime to the MBL regime. The Hamiltonian is written as
\begin{equation}
    H = \sum_j -t(a^\dagger_{j+1} a_j + a^\dagger_j a_{j+1}) + U_\infty n_jn_j + U n_{j}n_{j+1} + \mu_jn_j
    \label{eq:H}
\end{equation}
where $a^\dagger_j,a_j$ are the bosonic creation and annihilation operators on site $j$, $n_j = a^\dagger_j a_j$ is the density operator, $t$ is the hopping parameter, $U_\infty\rightarrow\infty$ imposes the hardcore constraint, $U$ is the nearest neighbor interaction strength, and $\mu_j$ is the disordered chemical potential on site $j$ taken from the uniform distribution $[-W,W]$. We fix $t=1$ throughout the calculations presented below. This model is known to exhibit a dome like phase boundary between the ergodic and MBL regimes as a function of $U$ and $W$~\cite{BarLev15,Kondov15} though the fate of MBL in the thermodynamic limit is still an open question.

\section{The Quantum Metric in Anderson Localization}
\begin{figure}
    \centering
    \includegraphics[width=\linewidth]{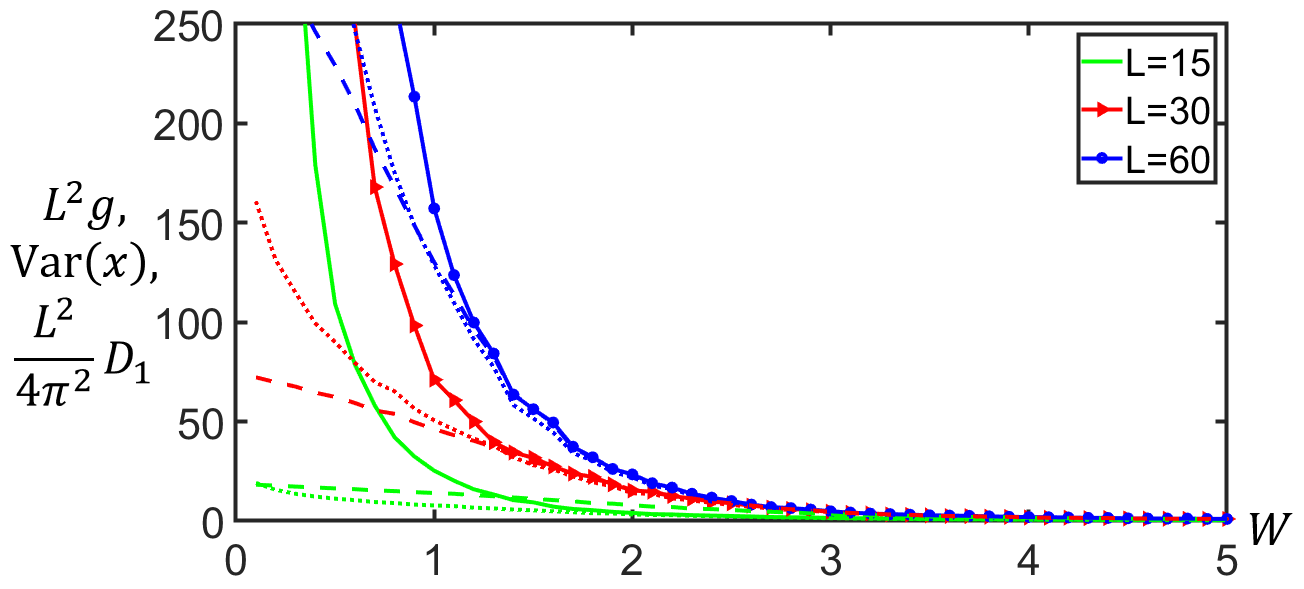}
    \caption{MBQM $L^2g$ (solid), variance (dashed) and $(L/2\pi)^2 D_1$ (dotted) as a function of disorder strength $W$ for a single particle Anderson Insulator system. The line color corresponds to different system lengths. These quantities are all equivalent in a localized state. The discrepancy at low disorder arises from the localization length becoming comparable to the system size.}
    \label{fig:AndersonInsulator}
\end{figure}

Before considering a many-body system, we demonstrate that this theoretical framework captures the localization properties of the single particle Anderson insulator, setting $U=0$ in Eq.~\ref{eq:H}. We calculate the MBQM and $D_1$ of the eigenstate in the middle of the energy spectrum, taking the median over 1000 disorder realizations for systems sizes ranging from $L=15$ to 60 for disorder parameter $W$ from 0.1 to 8. (We take the median as we find it converges more quickly than the average. The average converges more slowly as there are some disorder realizations where the quantum metric is very large due to the gapless nature of these excited states in the thermodynamic limit.) In one dimension, a disordered single particle system is expected to be an insulator for any disorder strength and thus we do not expect any metal-insulator transition. We present the values of $L^2 g$ (solid), $(L/2\pi)^2 D_1$ (dotted), and $\text{Var}(X)$ (dashed) as a function of disorder $W$ in Fig.~\ref{fig:AndersonInsulator}. The variance is calculated by shifting the localized state by its center-of-mass so that boundary effects are minimized and calculating as though it were an open boundary system where the typical position operator is well-defined. This calculation of the variance is only accurate when the shifted wave function does not cross the PBC.

In an Anderson insulating state, the localization length is expected to saturate to a finite value for a sufficiently large system. Additionally, as an insulating state, the MBQM and the localization parameter should approximately agree, converging as the system size increases. This is exactly what we observe for large disorder in Fig.~\ref{fig:AndersonInsulator}, where all three quantities agree and are independent of system size. For intermediate disorder, the three quantities still agree, but there is some noticeable dependence on system size suggesting that we have not achieved a large enough system to saturate the localization length. For low disorder, discrepancies between the MBQM, variance and localization parameter arise due to finite size effects since the localization length of the state becomes larger than the finite system, and thus the system's insulating character is hidden to the localization parameter. Evidence that this is a finite size effect is seen in how the agreement improves at lower disorder strengths as the system size increases. We emphasize that these results are obtained for an excited state where there is no energy gap in the thermodynamic limit. Still the numerical results show that the MBQM, the localization length and the variance agree when the disorder is sufficiently strong in finite systems. This provides evidence that in these disordered systems the MBQM is still meaningfully related to the localization length even though the system is gapless in the thermodynamic limit because in finite disordered systems there is a vanishingly small probability that two states are exactly degenerate.

We next demonstrate that this theoretical framework applies in a many-body setting by investigating the Anderson insulator at half-filling. Here, we still expect that the MBQM and localization parameter should agree for all disorder strengths and saturate to some finite value for a large enough system size. We will calculate the MBQM and $D_N$ at half-filling for disorder strengths up to $W=8$ at system lengths from $L=8$ to 18, taking the median of the MBQM and localization parameter over 1000 disorder realizations. We focus on the state in the middle of the spectrum corresponding to "infinite temperature" in the thermodynamic limit. Note in the thermodynamic limit this part of the spectrum is gapless. We implement an algorithm for calculation of the quantum metric that does not require full diagonalization of the Hamiltonian relying on the shift-and-invert method. An outline of the algorithm is presented in Table~\ref{tab:alg}. For the largest system, we are unable to obtain values when $W<1$ due to memory limitations.

\begin{table*}[]
    \centering
    \begin{tabular}{l}
        Algorithm \\
        \hline\hline
        \it{$H$ is the Hamiltonian}\\ 
        \it{numB is the size of the Hilbert space}\\
        \it{dHnu and dHmu are $\partial_{\nu} H$ and $\partial_{\mu} H$, respectively.}\\
        \hline\hline
        E  =  eigs( H, 10,'bothendsreal')\\
        Eshift  =  (max(E)+min(E))/2\\
        eigvec, eigval  =  eigs(H - Eshift*speye(numB),10,'smallestabs')\\
        E0, index0  =  min(abs(diag(eigval)))\\
        E0  =  eigval(index0,index0)\\
        phi0  =  eigvec(:,index0)\\
        dHnuPhi0  =  dHnu*phi0\\
        dHmuPhi0  =  dHmu*phi0\\
        HE2dHnuPhi0  =  mldivide(((H-(Eshift + E0)*speye(numB))*(H-(Eshift+E0)*speye(numB))),dHnuphi0)\\
        Qmunu  =  dHmuPhi0'*HE2dHnuPhi0\\
        \hline\hline
    \end{tabular}
    \caption{Algorithm for calculating the quantum metric without full diagonalization as implemented in MATLAB. This algorithm allows us to overcome system size limitations by diagonalizing only the extremal eigenvalue which can be obtained by various numerical methods such as Lanczos algorithm.  The trade-off is we have to perform this partial diagonalization twice and perform a matrix inversion.}
    \label{tab:alg}
\end{table*}

\begin{figure}
    \centering
    \includegraphics[width=\linewidth]{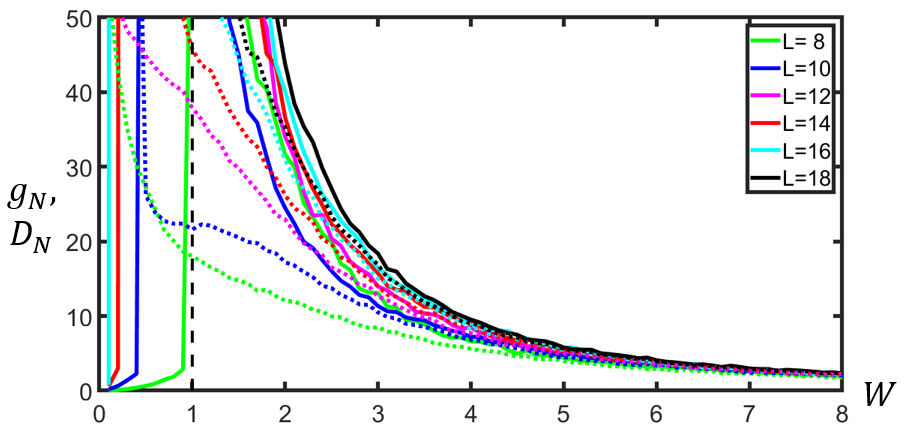}
    \caption{Quantum metric, $g_N$, and localization length parameter, $D_N$, as a function of disorder strength $W$ for a half-filled Anderson insulator. The colors correspond to different lengths of the system, $L$, as labeled in the legend. The key feature to notice is that for large enough disorder the two quantities agree and become independent of system size. Disagreements at low disorder are attributed to finite size effects.}
    \label{fig:MBAnderson}
\end{figure}

We present the results for the MBQM and the localization parameter in Fig.~\ref{fig:MBAnderson}. As in the single particle case, we observe that the MBQM and localization parameter agree for large disorder and for a wider range of disorder strengths with larger system size. For large disorder, the MBQM and localization parameter saturate, becoming independent of system size. The discrepancies at low disorder are again attributed to finite size effects. We emphasize again that the part of the spectrum we are considering is gapless in the thermodynamic limit. Still we find that the MBQM and the localization parameter are related in finite size systems.

\begin{figure}
    \centering
    \includegraphics[width=\linewidth]{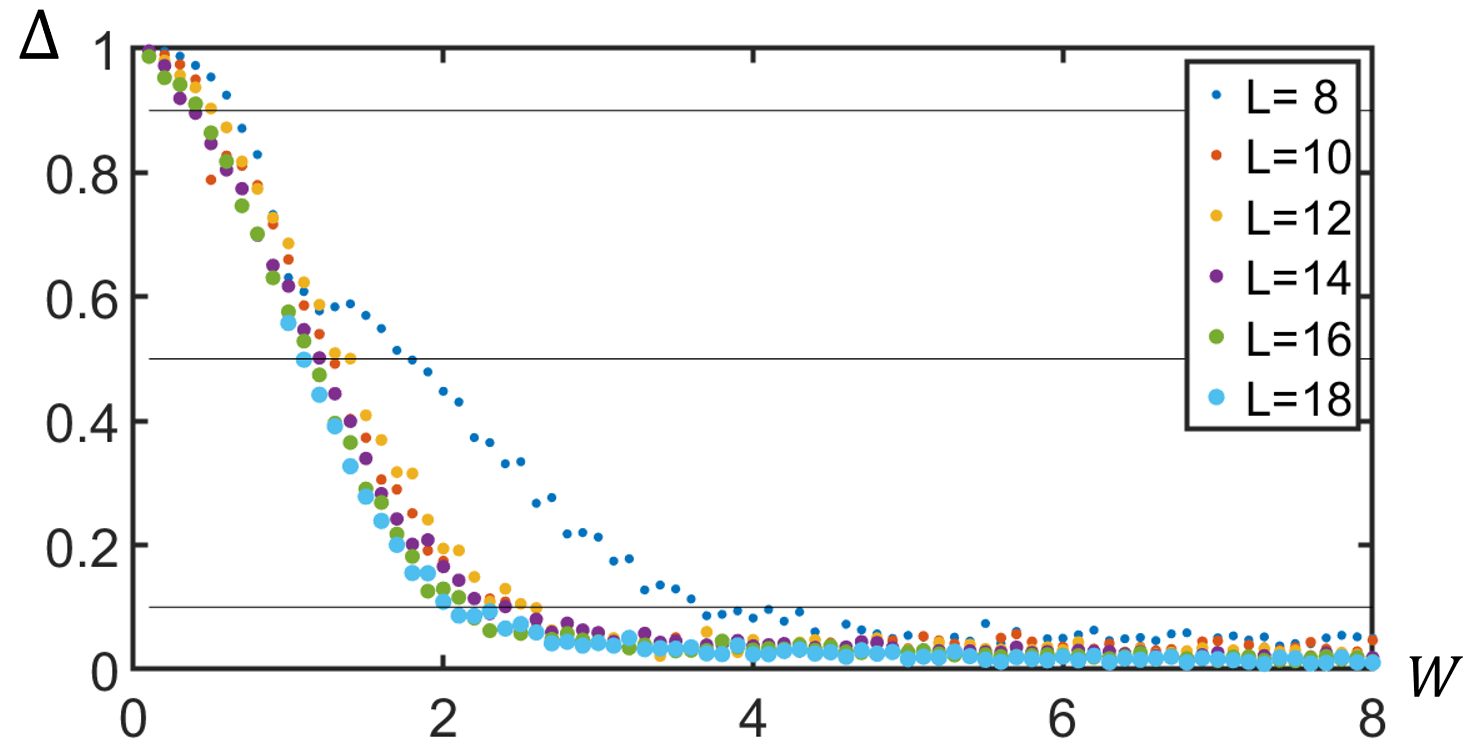}
    \caption{The parameter $\Delta$ for the half-filled non interacting disordered chain. The point size and color correspond to different system lengths, $L$, as labeled in the legend with larger points representing longer chains. For large disorder we see that the parameter drops to zero as we have reached a large enough system size to capture the insulating nature of the state.}
    \label{fig:DeltaAI}
\end{figure}

We introduce a quantity to measure the agreement between the MBQM and localization parameter. We define the quantity
\begin{equation}
    \Delta = \frac{|g_N-D_N|}{g_N+D_N}
\end{equation}
constructed such that it goes to 0 in an insulator where the MBQM and localization parameter agree and 1 in a conductor where the two quantities are unrelated and the localization parameter diverges. The values of this parameter are presented for the half-filled Anderson insulator in Fig.~\ref{fig:DeltaAI}. We observe that for large disorder the parameter is near zero while for small disorder the parameter moves to 1. We know that in this model the state is insulating for any disorder strength, and thus expect that the parameter should be zero across the whole plot. The discrepancy suggests that we have not achieved a large enough system size to obtain the localization properties for these low disorder strengths. Our $\Delta$ parameter has the advantage that we can determine whether our system appears insulating or conducting without performing a thermodynamic limit. In general, the localization parameter alone determines whether a state is conducting or insulating, but this requires one to obtain the correct thermodynamic limit. Unfortunately, a proper scaling for the localization parameter or MBQM for these excited states is not currently known. The $\Delta$ parameter allows us to establish the state as insulating despite the relatively small sizes available to us from exact diagonalization.

\section{Many-Body Localization}
Now we turn on the nearest-neighbor interaction to study the MBL transition. In contrast to the noninteracting case where the 1D disordered system is always localized, interacting disordered systems are known to exhibit a transition between an ergodic delocalized regime at low disorder and an MBL regime where ergodicity breaks down at large disorder. The critical disorder separating the two regimes depends on both the interaction strength and the system size. We calculate the MBQM and localization parameter at half-filling for a range of interaction and disorder strengths for system lengths from $L=8$ to 18. We take the median over 1000 disorder realizations for the eigenstate from the center of the spectrum, corresponding to the "infinite temperature" state.

\begin{figure*}
    \centering
    \includegraphics[width=\linewidth]{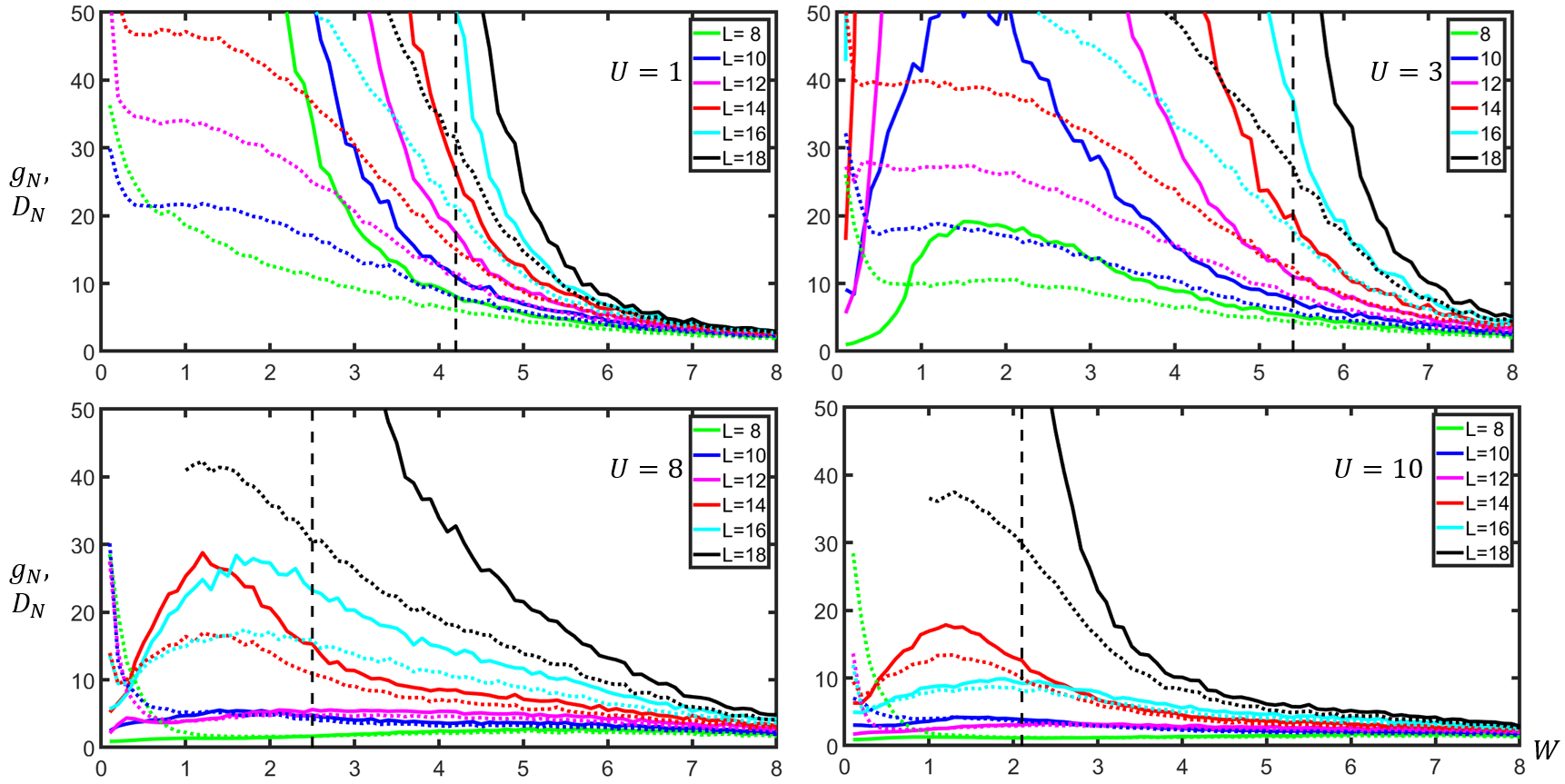}
    \caption{Many-body quantum metric $g_N=4\pi^2Ng$ (solid) and localization parameter $D_N=-N\ln|z_N|^2$ (dashed) versus disorder strength for different fixed interaction strengths $U$. The line color corresponds to different system lengths, $L$, as labeled in the legend. Vertical dashed lines correspond to the contour presented on the phase diagram on the right panel of Fig.~\ref{fig:PDfinite} where the parameter $\Delta=0.5$ for the system with $L=18$.}
    \label{fig:Geo_ZvsW}
\end{figure*}

We present the values of $g_N$ (solid) and $D_N$ (dashed) for each system size as a function of $W$ for several different values of the interaction strength $U$ in Fig.~\ref{fig:Geo_ZvsW}. We see again that for large disorder the MBQM and localization parameter agree while the system size dependence weakens. We interpret the agreement between the MBQM and the localization parameter as evidence that we are in an insulating MBL regime while the weak system size dependence suggests we have not yet achieved a large enough system size to saturate to the true localization length over much of the MBL regime. We also observe that the MBQM appears to be strongly diverging as the disorder is decreased, suggesting that the MBQM is divergent for these states in the ergodic regime.

\begin{figure*}
    \centering
    \includegraphics[width=\linewidth]{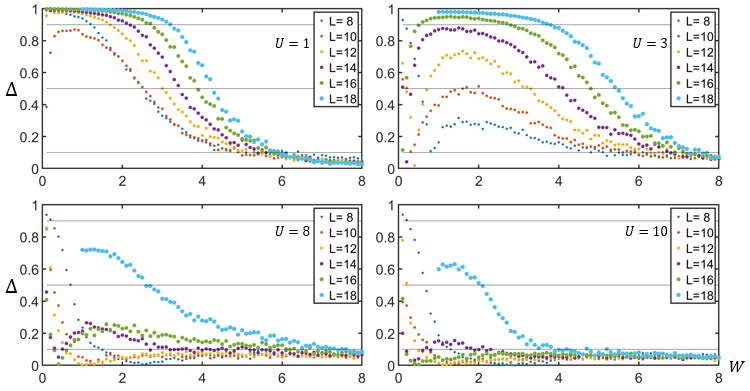}
    \caption{The parameter $\Delta$ as a function of disorder strength $W$ for different interaction strengths $U$. Each plot contains results for system sizes $L=8$ to $18$ with marker size proportional to system length.}
    \label{fig:DeltaFull}
\end{figure*}

Since neither the MBQM nor the localization parameter have saturated, we again employ the parameter $\Delta$ to determine whether we are in the ergodic or MBL regime. We plot the values of $\Delta$ as a function of disorder strength $W$ for different interaction strengths $U$ in Fig.~\ref{fig:DeltaFull} for system sizes $L=8$ to $18$. The behavior of $\Delta$ shows that we are indeed finding a transition between a conducting state at low disorder and an insulating state at large disorder with the critical disorder depending on the interaction strength as has been previously reported in studies of other indicators of the MBL transition. We also observe that as the interaction strength increases, the transition becomes less clear, suggesting that larger interaction strength requires larger system sizes to capture accurately.

\begin{figure}
    \centering
    \includegraphics[width=\linewidth]{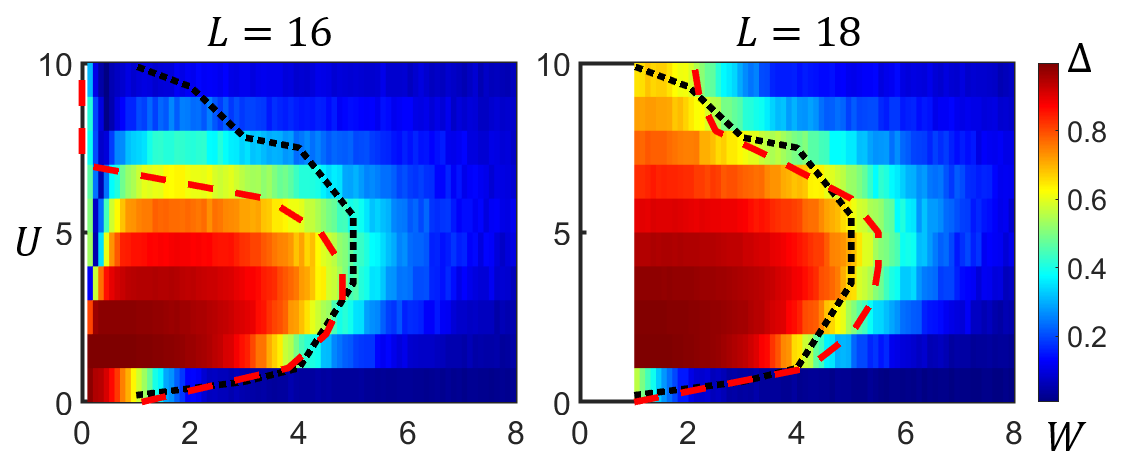}
    \caption{Phase diagram in the $W-U$ plane obtained from $\Delta$ for system sizes $L=16$ and $18$. The coloring is given by the value of $\Delta$. The black dotted curve is the phase boundary reported in Ref.~\cite{BarLev15}. The red dashed line is our phase boundary where $\Delta$ has dropped to 0.5.}
    \label{fig:PDfinite}
\end{figure}

We construct phase diagrams in the $W-U$ plane from the values of $\Delta$ for $L=16$ and $L=18$, presented in Fig.~\ref{fig:PDfinite}. We observe that the region where $\Delta$ is approximately 1, corresponding to the ergodic regime, forms a dome shape, as has been previously reported in studies of the MBL transition\cite{BarLev15,Serbyn15,Mondaini15,Chen23}. As a guide, we mark the contour where $\Delta$ drops to 0.5 as a suggestion of the approximate phase boundary. We also include the phase boundary reported in Ref.~\cite{BarLev15} obtained from analysis of the dynamical exponents, which agrees quite well with our proposed phase boundary for $L=18$. From this agreement, we are confident that our parameter $\Delta$ does in fact capture the MBL transition. Note that the phase boundaries are system size dependent and are expected to drift significantly as the system size increases, but we are unable to provide a meaningful extrapolation from the limited system sizes available.

\section{Localization length}
\begin{figure}
    \centering
    \includegraphics[width = \linewidth]{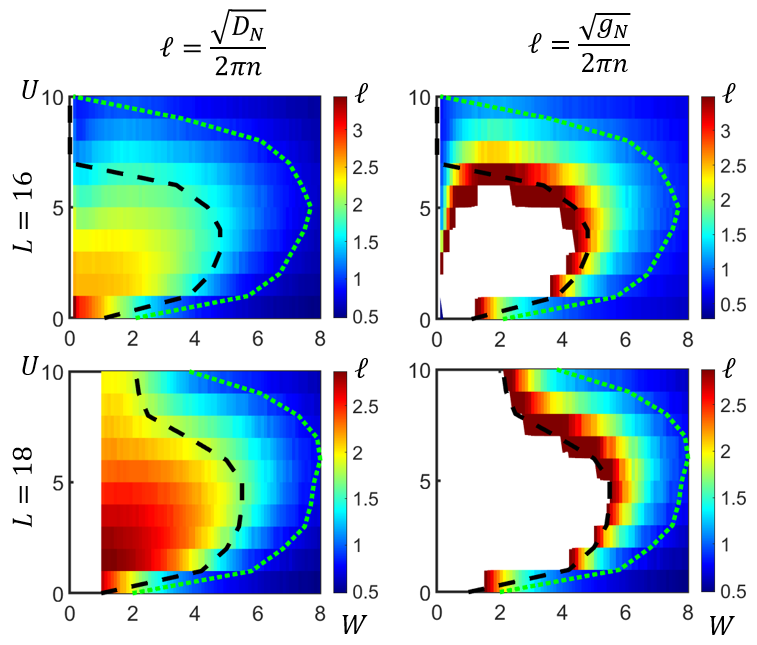}
    \caption{Localization length for system sizes $L=16$ and $L=18$. We present localization lengths extracted from both the localization parameter and the MBQM. The localization lengths presented are only meaningful in the MBL regime which occurs to the right of the dashed black phase boundary. Between the black dashed and the dotted green phase boundaries, the phase appears to be intermediate and not yet fully localized, likely the previously reported quantum critical phase.}
    \label{fig:LocLen}
\end{figure}

Having established that $\Delta$ captures the ergodic-MBL phase transition, we now present the extracted localization lengths from the MBQM and localization parameter. As stated earlier, in 1D the localization parameter and MBQM define a localization length as
\begin{equation}
    \ell_N=\frac{\sqrt{D_N}}{2\pi n} = \frac{\sqrt{g_N}}{ n}
\end{equation}
where $n=N/L$ is the density. We present the localization lengths for $L=16$ and $18$ in the $W-U$ plane in Fig.~\ref{fig:LocLen}. For each system size, we present localization lengths obtained from both the localization parameter (left) and the MBQM (right). The values presented can only meaningfully be interpreted as the localization length when the system is in a localized phase. We plot contours where $\Delta=0.5$ (black dashed) and $\Delta = 0.1$ (green dotted). From the figure we observe that in the region between dashed black and dotted green contours, the localization lengths obtained from the MBQM and localization parameter have some discrepancies with the MBQM based length being somewhat larger. This results from our definition that that the transition should occur for $\Delta=0.5$ and the fact that we are looking at the finite size system crossover between the two regimes. In previous studies of the MBL transition, this region has been observed and is a quantum critical phase that is expected to disappear in the thermodynamic limit. This is a limitation of our analysis and is a common issue for many finite size system studies of the MBL transition. As the system size increases, we expect that this quantum critical phase will shrink and the agreement between the two localization lengths will improve as the MBL regime stabilizes. 

\section{Conclusion}
In this work we have demonstrated that the transition from the ergodic regime to the MBL regime can be characterized by investigating the MBQM and the localization parameter together, taking their agreement as a signature of the onset of localization. This is surprising as their agreement holds even in highly excited states, which are not gapped in the thermodynamic limit. Still, for a finite system, there is generically no degeneracies in the excited states of a given disorder configuration. Thus, we can calculate the MBQM and observe qualitative differences in the behavior of the MBQM and localization parameter in the ergodic and MBL regimes. This analysis allows for a new definition of a localization length in the MBL regime. This localization length has the advantage that it is the natural localization length of insulating states as proposed by Resta~\cite{Resta99}, for example giving the magnetic length in quantum Hall insulators~\cite{Resta05}. The success of this quantity in capturing the transition is perhaps unsurprising as this transition should in principle be a metal-insulator transition. Note we do not expect our method to detect ergodic-MBL transitions that may occur as insulator-insulator transitions, an example of which has recently been reported~\cite{Wilkens23}. Furthermore, we note that MBQM is an experimentally measurable quantity. One method to measure MBQM is through the AC response~\cite{Ozawa19}, which was used in an ultracold atom experiment to extract the MBQM~\cite{Asteria2019}. Recently, it was pointed out that the MBQM is directly related to the small wavevector limit of the static structure factor~\cite{Onishi2025}; this relation was used to estimate the MBQM of LiF through inelastic X-ray scattering~\cite{Balut2025}. Our result, relating the localization length of MBL to the MBQM, thus provides a direct experimental means to measure the localization length of MBL. We add that preparing an initial highly excited eigenstate is in principle possible based on proposals utilizing local quenches~\cite{Liu23} in cold atom systems. It is exponentially difficult to construct an exact eigenstate, but preparation of an approximate eigenstate can be informed by advances in work with quantum hardware processors that have used LIOMs~\cite{Chandran15} or variational quantum eigensolver~\cite{Guo18} to approximate MBL eigenstates. Another possibility is to modify our calculation so that we consider the quantum metric associated with a specified energy window.

While at this stage we are unable to make definitive statements about the fate of MBL in the thermodynamic limit, there is potential for this analysis to shed light on this difficult problem as we push to larger system sizes with techniques beyond exact diagonalization. The results presented in Fig.~\ref{fig:Geo_ZvsW}, suggest that despite the system becoming gapless in the thermodynamic limit, the MBQM appears to be converging in the median. In order for this to hold in the thermodynamic limit, a careful analysis of Eq.~\ref{eq:gmunu2} is required. Its possible that while the energy difference goes to zero in the thermodynamic limit, so too does the numerator ($\langle\phi_0|\partial_\theta H|\phi_n\rangle$), canceling the divergence. Additionally, further pushing this analysis in MBL can help advance techniques in extracting the quantum metric as has been recently seen in the study of disordered topological insulators~\cite{Romeral24}. Another interesting avenue to explore is to develop a proper theory of the scaling behavior of the localization parameter, MBQM and parameter $\Delta$, which can potentially contribute to our understanding of the critical regime. This critical regime has been explored in efforts to determine if the MBL transition is a Berezinskii-Kosterlitz-Thouless transition~\cite{Dumitrescu19,Goremykina19} with some numerical evidence provided by the scaling behavior near the critical point~\cite{Aramthottil21}.

\section{Acknowledgements}
We thank Ryusuke Hamazaki and Henning Schomerus for discussions on many-body localization.
This work has been supported by JSPS KAKENHI Grant Number JP20H01845, JST PRESTO Grant No. JPMJPR2353, JST CREST Grant Number JPMJCR19T1.

\bibliography{biblio.bib}

\begin{thebibliography}{88}%
\makeatletter
\providecommand \@ifxundefined [1]{%
 \@ifx{#1\undefined}
}%
\providecommand \@ifnum [1]{%
 \ifnum #1\expandafter \@firstoftwo
 \else \expandafter \@secondoftwo
 \fi
}%
\providecommand \@ifx [1]{%
 \ifx #1\expandafter \@firstoftwo
 \else \expandafter \@secondoftwo
 \fi
}%
\providecommand \natexlab [1]{#1}%
\providecommand \enquote  [1]{``#1''}%
\providecommand \bibnamefont  [1]{#1}%
\providecommand \bibfnamefont [1]{#1}%
\providecommand \citenamefont [1]{#1}%
\providecommand \href@noop [0]{\@secondoftwo}%
\providecommand \href [0]{\begingroup \@sanitize@url \@href}%
\providecommand \@href[1]{\@@startlink{#1}\@@href}%
\providecommand \@@href[1]{\endgroup#1\@@endlink}%
\providecommand \@sanitize@url [0]{\catcode `\\12\catcode `\$12\catcode
  `\&12\catcode `\#12\catcode `\^12\catcode `\_12\catcode `\%12\relax}%
\providecommand \@@startlink[1]{}%
\providecommand \@@endlink[0]{}%
\providecommand \url  [0]{\begingroup\@sanitize@url \@url }%
\providecommand \@url [1]{\endgroup\@href {#1}{\urlprefix }}%
\providecommand \urlprefix  [0]{URL }%
\providecommand \Eprint [0]{\href }%
\providecommand \doibase [0]{https://doi.org/}%
\providecommand \selectlanguage [0]{\@gobble}%
\providecommand \bibinfo  [0]{\@secondoftwo}%
\providecommand \bibfield  [0]{\@secondoftwo}%
\providecommand \translation [1]{[#1]}%
\providecommand \BibitemOpen [0]{}%
\providecommand \bibitemStop [0]{}%
\providecommand \bibitemNoStop [0]{.\EOS\space}%
\providecommand \EOS [0]{\spacefactor3000\relax}%
\providecommand \BibitemShut  [1]{\csname bibitem#1\endcsname}%
\let\auto@bib@innerbib\@empty
\bibitem [{\citenamefont {Bloch}\ \emph {et~al.}(2008)\citenamefont {Bloch},
  \citenamefont {Dalibard},\ and\ \citenamefont {Zwerger}}]{Bloch08}%
  \BibitemOpen
  \bibfield  {author} {\bibinfo {author} {\bibfnamefont {I.}~\bibnamefont
  {Bloch}}, \bibinfo {author} {\bibfnamefont {J.}~\bibnamefont {Dalibard}},\
  and\ \bibinfo {author} {\bibfnamefont {W.}~\bibnamefont {Zwerger}},\
  }\bibfield  {title} {\bibinfo {title} {Many-body physics with ultracold
  gases},\ }\href {https://doi.org/10.1103/RevModPhys.80.885} {\bibfield
  {journal} {\bibinfo  {journal} {Rev. Mod. Phys.}\ }\textbf {\bibinfo {volume}
  {80}},\ \bibinfo {pages} {885} (\bibinfo {year} {2008})}\BibitemShut
  {NoStop}%
\bibitem [{\citenamefont {Bilokon}\ \emph {et~al.}(2023)\citenamefont
  {Bilokon}, \citenamefont {Bilokon}, \citenamefont {Bañuls}, \citenamefont
  {Cichy},\ and\ \citenamefont {Sotnikov}}]{Bilokon23}%
  \BibitemOpen
  \bibfield  {author} {\bibinfo {author} {\bibfnamefont {V.}~\bibnamefont
  {Bilokon}}, \bibinfo {author} {\bibfnamefont {E.}~\bibnamefont {Bilokon}},
  \bibinfo {author} {\bibfnamefont {M.~C.}\ \bibnamefont {Bañuls}}, \bibinfo
  {author} {\bibfnamefont {A.}~\bibnamefont {Cichy}},\ and\ \bibinfo {author}
  {\bibfnamefont {A.}~\bibnamefont {Sotnikov}},\ }\bibfield  {title} {\bibinfo
  {title} {Many-body correlations in one-dimensional optical lattices with
  alkaline-earth(-like) atoms},\ }\bibfield  {journal} {\bibinfo  {journal}
  {Scientific Reports}\ }\textbf {\bibinfo {volume} {13}},\ \href
  {https://doi.org/10.1038/s41598-023-37077-1} {10.1038/s41598-023-37077-1}
  (\bibinfo {year} {2023})\BibitemShut {NoStop}%
\bibitem [{\citenamefont {Chomaz}\ \emph {et~al.}(2022)\citenamefont {Chomaz},
  \citenamefont {Ferrier-Barbut}, \citenamefont {Ferlaino}, \citenamefont
  {Laburthe-Tolra}, \citenamefont {Lev},\ and\ \citenamefont
  {Pfau}}]{Chomaz23}%
  \BibitemOpen
  \bibfield  {author} {\bibinfo {author} {\bibfnamefont {L.}~\bibnamefont
  {Chomaz}}, \bibinfo {author} {\bibfnamefont {I.}~\bibnamefont
  {Ferrier-Barbut}}, \bibinfo {author} {\bibfnamefont {F.}~\bibnamefont
  {Ferlaino}}, \bibinfo {author} {\bibfnamefont {B.}~\bibnamefont
  {Laburthe-Tolra}}, \bibinfo {author} {\bibfnamefont {B.~L.}\ \bibnamefont
  {Lev}},\ and\ \bibinfo {author} {\bibfnamefont {T.}~\bibnamefont {Pfau}},\
  }\bibfield  {title} {\bibinfo {title} {Dipolar physics: a review of
  experiments with magnetic quantum gases},\ }\href
  {https://doi.org/10.1088/1361-6633/aca814} {\bibfield  {journal} {\bibinfo
  {journal} {Reports on Progress in Physics}\ }\textbf {\bibinfo {volume}
  {86}},\ \bibinfo {pages} {026401} (\bibinfo {year} {2022})}\BibitemShut
  {NoStop}%
\bibitem [{\citenamefont {Malz}\ and\ \citenamefont {Cirac}(2023)}]{Malz23}%
  \BibitemOpen
  \bibfield  {author} {\bibinfo {author} {\bibfnamefont {D.}~\bibnamefont
  {Malz}}\ and\ \bibinfo {author} {\bibfnamefont {J.~I.}\ \bibnamefont
  {Cirac}},\ }\bibfield  {title} {\bibinfo {title} {Few-body analog quantum
  simulation with rydberg-dressed atoms in optical lattices},\ }\href
  {https://doi.org/10.1103/PRXQuantum.4.020301} {\bibfield  {journal} {\bibinfo
   {journal} {PRX Quantum}\ }\textbf {\bibinfo {volume} {4}},\ \bibinfo {pages}
  {020301} (\bibinfo {year} {2023})}\BibitemShut {NoStop}%
\bibitem [{\citenamefont {Blatt}\ and\ \citenamefont {Roos}(2012)}]{Blatt12}%
  \BibitemOpen
  \bibfield  {author} {\bibinfo {author} {\bibfnamefont {R.}~\bibnamefont
  {Blatt}}\ and\ \bibinfo {author} {\bibfnamefont {C.~F.}\ \bibnamefont
  {Roos}},\ }\bibfield  {title} {\bibinfo {title} {Quantum simulations with
  trapped ions},\ }\bibfield  {journal} {\bibinfo  {journal} {Nature Physics}\
  }\textbf {\bibinfo {volume} {8}},\ \href {https://doi.org/10.1038/nphys2252}
  {10.1038/nphys2252} (\bibinfo {year} {2012})\BibitemShut {NoStop}%
\bibitem [{\citenamefont {Adams}\ \emph {et~al.}(2019)\citenamefont {Adams},
  \citenamefont {Pritchard},\ and\ \citenamefont {Shaffer}}]{Adams20}%
  \BibitemOpen
  \bibfield  {author} {\bibinfo {author} {\bibfnamefont {C.~S.}\ \bibnamefont
  {Adams}}, \bibinfo {author} {\bibfnamefont {J.~D.}\ \bibnamefont
  {Pritchard}},\ and\ \bibinfo {author} {\bibfnamefont {J.~P.}\ \bibnamefont
  {Shaffer}},\ }\bibfield  {title} {\bibinfo {title} {Rydberg atom quantum
  technologies},\ }\href {https://doi.org/10.1088/1361-6455/ab52ef} {\bibfield
  {journal} {\bibinfo  {journal} {Journal of Physics B: Atomic, Molecular and
  Optical Physics}\ }\textbf {\bibinfo {volume} {53}},\ \bibinfo {pages}
  {012002} (\bibinfo {year} {2019})}\BibitemShut {NoStop}%
\bibitem [{\citenamefont {Wu}\ \emph {et~al.}(2021)\citenamefont {Wu},
  \citenamefont {Liang}, \citenamefont {Tian}, \citenamefont {Yang},
  \citenamefont {Chen}, \citenamefont {Liu}, \citenamefont {Tey},\ and\
  \citenamefont {You}}]{Wu21}%
  \BibitemOpen
  \bibfield  {author} {\bibinfo {author} {\bibfnamefont {X.}~\bibnamefont
  {Wu}}, \bibinfo {author} {\bibfnamefont {X.}~\bibnamefont {Liang}}, \bibinfo
  {author} {\bibfnamefont {Y.}~\bibnamefont {Tian}}, \bibinfo {author}
  {\bibfnamefont {F.}~\bibnamefont {Yang}}, \bibinfo {author} {\bibfnamefont
  {C.}~\bibnamefont {Chen}}, \bibinfo {author} {\bibfnamefont {Y.-C.}\
  \bibnamefont {Liu}}, \bibinfo {author} {\bibfnamefont {M.~K.}\ \bibnamefont
  {Tey}},\ and\ \bibinfo {author} {\bibfnamefont {L.}~\bibnamefont {You}},\
  }\bibfield  {title} {\bibinfo {title} {A concise review of rydberg atom based
  quantum computation and quantum simulation*},\ }\href
  {https://doi.org/10.1088/1674-1056/abd76f} {\bibfield  {journal} {\bibinfo
  {journal} {Chinese Physics B}\ }\textbf {\bibinfo {volume} {30}},\ \bibinfo
  {pages} {020305} (\bibinfo {year} {2021})}\BibitemShut {NoStop}%
\bibitem [{\citenamefont {Kinoshita}\ \emph {et~al.}(2006)\citenamefont
  {Kinoshita}, \citenamefont {Wenger},\ and\ \citenamefont
  {Weiss}}]{Kinoshita06}%
  \BibitemOpen
  \bibfield  {author} {\bibinfo {author} {\bibfnamefont {T.}~\bibnamefont
  {Kinoshita}}, \bibinfo {author} {\bibfnamefont {T.}~\bibnamefont {Wenger}},\
  and\ \bibinfo {author} {\bibfnamefont {D.~S.}\ \bibnamefont {Weiss}},\
  }\bibfield  {title} {\bibinfo {title} {A quantum newton's cradle},\
  }\bibfield  {journal} {\bibinfo  {journal} {Nature}\ }\textbf {\bibinfo
  {volume} {440}},\ \href {https://doi.org/10.1038/nature04693}
  {10.1038/nature04693} (\bibinfo {year} {2006})\BibitemShut {NoStop}%
\bibitem [{\citenamefont {Vasseur}\ and\ \citenamefont
  {Moore}(2016)}]{Vasseur16}%
  \BibitemOpen
  \bibfield  {author} {\bibinfo {author} {\bibfnamefont {R.}~\bibnamefont
  {Vasseur}}\ and\ \bibinfo {author} {\bibfnamefont {J.~E.}\ \bibnamefont
  {Moore}},\ }\bibfield  {title} {\bibinfo {title} {Nonequilibrium quantum
  dynamics and transport: from integrability to many-body localization},\
  }\href {https://doi.org/10.1088/1742-5468/2016/06/064010} {\bibfield
  {journal} {\bibinfo  {journal} {Journal of Statistical Mechanics: Theory and
  Experiment}\ }\textbf {\bibinfo {volume} {2016}},\ \bibinfo {pages} {064010}
  (\bibinfo {year} {2016})}\BibitemShut {NoStop}%
\bibitem [{\citenamefont {Nandkishore}\ and\ \citenamefont
  {Huse}(2015)}]{Nandkishore15}%
  \BibitemOpen
  \bibfield  {author} {\bibinfo {author} {\bibfnamefont {R.}~\bibnamefont
  {Nandkishore}}\ and\ \bibinfo {author} {\bibfnamefont {D.~A.}\ \bibnamefont
  {Huse}},\ }\bibfield  {title} {\bibinfo {title} {Many-body localization and
  thermalization in quantum statistical mechanics},\ }\href
  {https://doi.org/10.1146/annurev-conmatphys-031214-014726} {\bibfield
  {journal} {\bibinfo  {journal} {Annual Review of Condensed Matter Physics}\
  }\textbf {\bibinfo {volume} {6}},\ \bibinfo {pages} {15} (\bibinfo {year}
  {2015})},\ \Eprint
  {https://arxiv.org/abs/https://doi.org/10.1146/annurev-conmatphys-031214-014726}
  {https://doi.org/10.1146/annurev-conmatphys-031214-014726} \BibitemShut
  {NoStop}%
\bibitem [{\citenamefont {Luca}\ and\ \citenamefont
  {Scardicchio}(2013)}]{DeLuca13}%
  \BibitemOpen
  \bibfield  {author} {\bibinfo {author} {\bibfnamefont {A.~D.}\ \bibnamefont
  {Luca}}\ and\ \bibinfo {author} {\bibfnamefont {A.}~\bibnamefont
  {Scardicchio}},\ }\bibfield  {title} {\bibinfo {title} {Ergodicity breaking
  in a model showing many-body localization},\ }\href
  {https://doi.org/10.1209/0295-5075/101/37003} {\bibfield  {journal} {\bibinfo
   {journal} {Europhysics Letters}\ }\textbf {\bibinfo {volume} {101}},\
  \bibinfo {pages} {37003} (\bibinfo {year} {2013})}\BibitemShut {NoStop}%
\bibitem [{\citenamefont {Levi}\ \emph {et~al.}(2016)\citenamefont {Levi},
  \citenamefont {Heyl}, \citenamefont {Lesanovsky},\ and\ \citenamefont
  {Garrahan}}]{Levi16}%
  \BibitemOpen
  \bibfield  {author} {\bibinfo {author} {\bibfnamefont {E.}~\bibnamefont
  {Levi}}, \bibinfo {author} {\bibfnamefont {M.}~\bibnamefont {Heyl}}, \bibinfo
  {author} {\bibfnamefont {I.}~\bibnamefont {Lesanovsky}},\ and\ \bibinfo
  {author} {\bibfnamefont {J.~P.}\ \bibnamefont {Garrahan}},\ }\bibfield
  {title} {\bibinfo {title} {Robustness of many-body localization in the
  presence of dissipation},\ }\href
  {https://doi.org/10.1103/PhysRevLett.116.237203} {\bibfield  {journal}
  {\bibinfo  {journal} {Phys. Rev. Lett.}\ }\textbf {\bibinfo {volume} {116}},\
  \bibinfo {pages} {237203} (\bibinfo {year} {2016})}\BibitemShut {NoStop}%
\bibitem [{\citenamefont {Luca~D'Alessio}\ and\ \citenamefont
  {Rigol}(2016)}]{DAlessio16}%
  \BibitemOpen
  \bibfield  {author} {\bibinfo {author} {\bibfnamefont {A.~P.}\ \bibnamefont
  {Luca~D'Alessio}, \bibfnamefont {Yariv~Kafri}}\ and\ \bibinfo {author}
  {\bibfnamefont {M.}~\bibnamefont {Rigol}},\ }\bibfield  {title} {\bibinfo
  {title} {From quantum chaos and eigenstate thermalization to statistical
  mechanics and thermodynamics},\ }\href
  {https://doi.org/10.1080/00018732.2016.1198134} {\bibfield  {journal}
  {\bibinfo  {journal} {Advances in Physics}\ }\textbf {\bibinfo {volume}
  {65}},\ \bibinfo {pages} {239} (\bibinfo {year} {2016})},\ \Eprint
  {https://arxiv.org/abs/https://doi.org/10.1080/00018732.2016.1198134}
  {https://doi.org/10.1080/00018732.2016.1198134} \BibitemShut {NoStop}%
\bibitem [{\citenamefont {Turner}\ \emph {et~al.}(2018)\citenamefont {Turner},
  \citenamefont {Michailidis}, \citenamefont {Abanin}, \citenamefont {Serbyn},\
  and\ \citenamefont {Papi\ifmmode~\acute{c}\else \'{c}\fi{}}}]{Turner18}%
  \BibitemOpen
  \bibfield  {author} {\bibinfo {author} {\bibfnamefont {C.~J.}\ \bibnamefont
  {Turner}}, \bibinfo {author} {\bibfnamefont {A.~A.}\ \bibnamefont
  {Michailidis}}, \bibinfo {author} {\bibfnamefont {D.~A.}\ \bibnamefont
  {Abanin}}, \bibinfo {author} {\bibfnamefont {M.}~\bibnamefont {Serbyn}},\
  and\ \bibinfo {author} {\bibfnamefont {Z.}~\bibnamefont
  {Papi\ifmmode~\acute{c}\else \'{c}\fi{}}},\ }\bibfield  {title} {\bibinfo
  {title} {Quantum scarred eigenstates in a rydberg atom chain: Entanglement,
  breakdown of thermalization, and stability to perturbations},\ }\href
  {https://doi.org/10.1103/PhysRevB.98.155134} {\bibfield  {journal} {\bibinfo
  {journal} {Phys. Rev. B}\ }\textbf {\bibinfo {volume} {98}},\ \bibinfo
  {pages} {155134} (\bibinfo {year} {2018})}\BibitemShut {NoStop}%
\bibitem [{\citenamefont {Anderson}(1958)}]{Anderson58}%
  \BibitemOpen
  \bibfield  {author} {\bibinfo {author} {\bibfnamefont {P.~W.}\ \bibnamefont
  {Anderson}},\ }\bibfield  {title} {\bibinfo {title} {Absence of diffusion in
  certain random lattices},\ }\href {https://doi.org/10.1103/PhysRev.109.1492}
  {\bibfield  {journal} {\bibinfo  {journal} {Phys. Rev.}\ }\textbf {\bibinfo
  {volume} {109}},\ \bibinfo {pages} {1492} (\bibinfo {year}
  {1958})}\BibitemShut {NoStop}%
\bibitem [{\citenamefont {Fleishman}\ and\ \citenamefont
  {Anderson}(1980)}]{Fleishman80}%
  \BibitemOpen
  \bibfield  {author} {\bibinfo {author} {\bibfnamefont {L.}~\bibnamefont
  {Fleishman}}\ and\ \bibinfo {author} {\bibfnamefont {P.~W.}\ \bibnamefont
  {Anderson}},\ }\bibfield  {title} {\bibinfo {title} {Interactions and the
  anderson transition},\ }\href {https://doi.org/10.1103/PhysRevB.21.2366}
  {\bibfield  {journal} {\bibinfo  {journal} {Phys. Rev. B}\ }\textbf {\bibinfo
  {volume} {21}},\ \bibinfo {pages} {2366} (\bibinfo {year}
  {1980})}\BibitemShut {NoStop}%
\bibitem [{\citenamefont {{Finkel'shtein}}(1983)}]{Finkelstein83}%
  \BibitemOpen
  \bibfield  {author} {\bibinfo {author} {\bibfnamefont {A.~M.}\ \bibnamefont
  {{Finkel'shtein}}},\ }\bibfield  {title} {\bibinfo {title} {{Influence of
  Coulomb interaction on the properties of disordered metals}},\ }\href@noop {}
  {\bibfield  {journal} {\bibinfo  {journal} {Soviet Journal of Experimental
  and Theoretical Physics}\ }\textbf {\bibinfo {volume} {57}},\ \bibinfo
  {pages} {97} (\bibinfo {year} {1983})}\BibitemShut {NoStop}%
\bibitem [{\citenamefont {Giamarchi}\ and\ \citenamefont
  {Schulz}(1988)}]{Giamarchi88}%
  \BibitemOpen
  \bibfield  {author} {\bibinfo {author} {\bibfnamefont {T.}~\bibnamefont
  {Giamarchi}}\ and\ \bibinfo {author} {\bibfnamefont {H.~J.}\ \bibnamefont
  {Schulz}},\ }\bibfield  {title} {\bibinfo {title} {Anderson localization and
  interactions in one-dimensional metals},\ }\href
  {https://doi.org/10.1103/PhysRevB.37.325} {\bibfield  {journal} {\bibinfo
  {journal} {Phys. Rev. B}\ }\textbf {\bibinfo {volume} {37}},\ \bibinfo
  {pages} {325} (\bibinfo {year} {1988})}\BibitemShut {NoStop}%
\bibitem [{\citenamefont {Altshuler}\ \emph {et~al.}(1997)\citenamefont
  {Altshuler}, \citenamefont {Gefen}, \citenamefont {Kamenev},\ and\
  \citenamefont {Levitov}}]{Altshuler97}%
  \BibitemOpen
  \bibfield  {author} {\bibinfo {author} {\bibfnamefont {B.~L.}\ \bibnamefont
  {Altshuler}}, \bibinfo {author} {\bibfnamefont {Y.}~\bibnamefont {Gefen}},
  \bibinfo {author} {\bibfnamefont {A.}~\bibnamefont {Kamenev}},\ and\ \bibinfo
  {author} {\bibfnamefont {L.~S.}\ \bibnamefont {Levitov}},\ }\bibfield
  {title} {\bibinfo {title} {Quasiparticle lifetime in a finite system: A
  nonperturbative approach},\ }\href
  {https://doi.org/10.1103/PhysRevLett.78.2803} {\bibfield  {journal} {\bibinfo
   {journal} {Phys. Rev. Lett.}\ }\textbf {\bibinfo {volume} {78}},\ \bibinfo
  {pages} {2803} (\bibinfo {year} {1997})}\BibitemShut {NoStop}%
\bibitem [{\citenamefont {Gornyi}\ \emph {et~al.}(2005)\citenamefont {Gornyi},
  \citenamefont {Mirlin},\ and\ \citenamefont {Polyakov}}]{Gornyi05}%
  \BibitemOpen
  \bibfield  {author} {\bibinfo {author} {\bibfnamefont {I.~V.}\ \bibnamefont
  {Gornyi}}, \bibinfo {author} {\bibfnamefont {A.~D.}\ \bibnamefont {Mirlin}},\
  and\ \bibinfo {author} {\bibfnamefont {D.~G.}\ \bibnamefont {Polyakov}},\
  }\bibfield  {title} {\bibinfo {title} {Interacting electrons in disordered
  wires: Anderson localization and low-$t$ transport},\ }\href
  {https://doi.org/10.1103/PhysRevLett.95.206603} {\bibfield  {journal}
  {\bibinfo  {journal} {Phys. Rev. Lett.}\ }\textbf {\bibinfo {volume} {95}},\
  \bibinfo {pages} {206603} (\bibinfo {year} {2005})}\BibitemShut {NoStop}%
\bibitem [{\citenamefont {Basko}\ \emph {et~al.}(2006)\citenamefont {Basko},
  \citenamefont {Aleiner},\ and\ \citenamefont {Altshuler}}]{Basko06}%
  \BibitemOpen
  \bibfield  {author} {\bibinfo {author} {\bibfnamefont {D.}~\bibnamefont
  {Basko}}, \bibinfo {author} {\bibfnamefont {I.}~\bibnamefont {Aleiner}},\
  and\ \bibinfo {author} {\bibfnamefont {B.}~\bibnamefont {Altshuler}},\
  }\bibfield  {title} {\bibinfo {title} {Metal–insulator transition in a
  weakly interacting many-electron system with localized single-particle
  states},\ }\href {https://doi.org/https://doi.org/10.1016/j.aop.2005.11.014}
  {\bibfield  {journal} {\bibinfo  {journal} {Annals of Physics}\ }\textbf
  {\bibinfo {volume} {321}},\ \bibinfo {pages} {1126} (\bibinfo {year}
  {2006})}\BibitemShut {NoStop}%
\bibitem [{\citenamefont {Bauer}\ and\ \citenamefont {Nayak}(2013)}]{Bauer13}%
  \BibitemOpen
  \bibfield  {author} {\bibinfo {author} {\bibfnamefont {B.}~\bibnamefont
  {Bauer}}\ and\ \bibinfo {author} {\bibfnamefont {C.}~\bibnamefont {Nayak}},\
  }\bibfield  {title} {\bibinfo {title} {Area laws in a many-body localized
  state and its implications for topological order},\ }\href
  {https://doi.org/10.1088/1742-5468/2013/09/P09005} {\bibfield  {journal}
  {\bibinfo  {journal} {Journal of Statistical Mechanics: Theory and
  Experiment}\ }\textbf {\bibinfo {volume} {2013}},\ \bibinfo {pages} {P09005}
  (\bibinfo {year} {2013})}\BibitemShut {NoStop}%
\bibitem [{\citenamefont {Serbyn}\ \emph
  {et~al.}(2013{\natexlab{a}})\citenamefont {Serbyn}, \citenamefont
  {Papi\ifmmode~\acute{c}\else \'{c}\fi{}},\ and\ \citenamefont
  {Abanin}}]{Serbyn13}%
  \BibitemOpen
  \bibfield  {author} {\bibinfo {author} {\bibfnamefont {M.}~\bibnamefont
  {Serbyn}}, \bibinfo {author} {\bibfnamefont {Z.}~\bibnamefont
  {Papi\ifmmode~\acute{c}\else \'{c}\fi{}}},\ and\ \bibinfo {author}
  {\bibfnamefont {D.~A.}\ \bibnamefont {Abanin}},\ }\bibfield  {title}
  {\bibinfo {title} {Universal slow growth of entanglement in interacting
  strongly disordered systems},\ }\href
  {https://doi.org/10.1103/PhysRevLett.110.260601} {\bibfield  {journal}
  {\bibinfo  {journal} {Phys. Rev. Lett.}\ }\textbf {\bibinfo {volume} {110}},\
  \bibinfo {pages} {260601} (\bibinfo {year} {2013}{\natexlab{a}})}\BibitemShut
  {NoStop}%
\bibitem [{\citenamefont {Huang}(2021)}]{Huang21}%
  \BibitemOpen
  \bibfield  {author} {\bibinfo {author} {\bibfnamefont {Y.}~\bibnamefont
  {Huang}},\ }\bibfield  {title} {\bibinfo {title} {Extensive entropy from
  unitary evolution},\ }\bibfield  {journal} {\bibinfo  {journal} {Preprints}\
  }\href {https://doi.org/10.20944/preprints202104.0254.v1}
  {10.20944/preprints202104.0254.v1} (\bibinfo {year} {2021})\BibitemShut
  {NoStop}%
\bibitem [{\citenamefont {Abanin}\ \emph {et~al.}(2019)\citenamefont {Abanin},
  \citenamefont {Altman}, \citenamefont {Bloch},\ and\ \citenamefont
  {Serbyn}}]{Abanin19}%
  \BibitemOpen
  \bibfield  {author} {\bibinfo {author} {\bibfnamefont {D.~A.}\ \bibnamefont
  {Abanin}}, \bibinfo {author} {\bibfnamefont {E.}~\bibnamefont {Altman}},
  \bibinfo {author} {\bibfnamefont {I.}~\bibnamefont {Bloch}},\ and\ \bibinfo
  {author} {\bibfnamefont {M.}~\bibnamefont {Serbyn}},\ }\bibfield  {title}
  {\bibinfo {title} {Colloquium: Many-body localization, thermalization, and
  entanglement},\ }\href {https://doi.org/10.1103/RevModPhys.91.021001}
  {\bibfield  {journal} {\bibinfo  {journal} {Rev. Mod. Phys.}\ }\textbf
  {\bibinfo {volume} {91}},\ \bibinfo {pages} {021001} (\bibinfo {year}
  {2019})}\BibitemShut {NoStop}%
\bibitem [{\citenamefont {Sierant}\ and\ \citenamefont
  {Zakrzewski}(2018)}]{Sierant18}%
  \BibitemOpen
  \bibfield  {author} {\bibinfo {author} {\bibfnamefont {P.}~\bibnamefont
  {Sierant}}\ and\ \bibinfo {author} {\bibfnamefont {J.}~\bibnamefont
  {Zakrzewski}},\ }\bibfield  {title} {\bibinfo {title} {Many-body localization
  of bosons in optical lattices},\ }\href
  {https://doi.org/10.1088/1367-2630/aabb17} {\bibfield  {journal} {\bibinfo
  {journal} {New Journal of Physics}\ }\textbf {\bibinfo {volume} {20}},\
  \bibinfo {pages} {043032} (\bibinfo {year} {2018})}\BibitemShut {NoStop}%
\bibitem [{\citenamefont {Lukin}\ \emph {et~al.}(2019)\citenamefont {Lukin},
  \citenamefont {Rispoli}, \citenamefont {Schittko}, \citenamefont {Tai},
  \citenamefont {Kaufman}, \citenamefont {Choi}, \citenamefont {Khemani},
  \citenamefont {Léonard},\ and\ \citenamefont {Greiner}}]{Lukin19}%
  \BibitemOpen
  \bibfield  {author} {\bibinfo {author} {\bibfnamefont {A.}~\bibnamefont
  {Lukin}}, \bibinfo {author} {\bibfnamefont {M.}~\bibnamefont {Rispoli}},
  \bibinfo {author} {\bibfnamefont {R.}~\bibnamefont {Schittko}}, \bibinfo
  {author} {\bibfnamefont {M.~E.}\ \bibnamefont {Tai}}, \bibinfo {author}
  {\bibfnamefont {A.~M.}\ \bibnamefont {Kaufman}}, \bibinfo {author}
  {\bibfnamefont {S.}~\bibnamefont {Choi}}, \bibinfo {author} {\bibfnamefont
  {V.}~\bibnamefont {Khemani}}, \bibinfo {author} {\bibfnamefont
  {J.}~\bibnamefont {Léonard}},\ and\ \bibinfo {author} {\bibfnamefont
  {M.}~\bibnamefont {Greiner}},\ }\bibfield  {title} {\bibinfo {title} {Probing
  entanglement in a many-body\&\#x2013;localized system},\ }\href
  {https://doi.org/10.1126/science.aau0818} {\bibfield  {journal} {\bibinfo
  {journal} {Science}\ }\textbf {\bibinfo {volume} {364}},\ \bibinfo {pages}
  {256} (\bibinfo {year} {2019})},\ \Eprint
  {https://arxiv.org/abs/https://www.science.org/doi/pdf/10.1126/science.aau0818}
  {https://www.science.org/doi/pdf/10.1126/science.aau0818} \BibitemShut
  {NoStop}%
\bibitem [{\citenamefont {Bar~Lev}\ \emph {et~al.}(2015)\citenamefont
  {Bar~Lev}, \citenamefont {Cohen},\ and\ \citenamefont {Reichman}}]{BarLev15}%
  \BibitemOpen
  \bibfield  {author} {\bibinfo {author} {\bibfnamefont {Y.}~\bibnamefont
  {Bar~Lev}}, \bibinfo {author} {\bibfnamefont {G.}~\bibnamefont {Cohen}},\
  and\ \bibinfo {author} {\bibfnamefont {D.~R.}\ \bibnamefont {Reichman}},\
  }\bibfield  {title} {\bibinfo {title} {Absence of diffusion in an interacting
  system of spinless fermions on a one-dimensional disordered lattice},\ }\href
  {https://doi.org/10.1103/PhysRevLett.114.100601} {\bibfield  {journal}
  {\bibinfo  {journal} {Phys. Rev. Lett.}\ }\textbf {\bibinfo {volume} {114}},\
  \bibinfo {pages} {100601} (\bibinfo {year} {2015})}\BibitemShut {NoStop}%
\bibitem [{\citenamefont {Schreiber}\ \emph {et~al.}(2015)\citenamefont
  {Schreiber}, \citenamefont {Hodgman}, \citenamefont {Bordia}, \citenamefont
  {Lüschen}, \citenamefont {Fischer}, \citenamefont {Vosk}, \citenamefont
  {Altman}, \citenamefont {Schneider},\ and\ \citenamefont
  {Bloch}}]{Schreiber15}%
  \BibitemOpen
  \bibfield  {author} {\bibinfo {author} {\bibfnamefont {M.}~\bibnamefont
  {Schreiber}}, \bibinfo {author} {\bibfnamefont {S.~S.}\ \bibnamefont
  {Hodgman}}, \bibinfo {author} {\bibfnamefont {P.}~\bibnamefont {Bordia}},
  \bibinfo {author} {\bibfnamefont {H.~P.}\ \bibnamefont {Lüschen}}, \bibinfo
  {author} {\bibfnamefont {M.~H.}\ \bibnamefont {Fischer}}, \bibinfo {author}
  {\bibfnamefont {R.}~\bibnamefont {Vosk}}, \bibinfo {author} {\bibfnamefont
  {E.}~\bibnamefont {Altman}}, \bibinfo {author} {\bibfnamefont
  {U.}~\bibnamefont {Schneider}},\ and\ \bibinfo {author} {\bibfnamefont
  {I.}~\bibnamefont {Bloch}},\ }\bibfield  {title} {\bibinfo {title}
  {Observation of many-body localization of interacting fermions in a
  quasirandom optical lattice},\ }\href
  {https://doi.org/10.1126/science.aaa7432} {\bibfield  {journal} {\bibinfo
  {journal} {Science}\ }\textbf {\bibinfo {volume} {349}},\ \bibinfo {pages}
  {842} (\bibinfo {year} {2015})},\ \Eprint
  {https://arxiv.org/abs/https://www.science.org/doi/pdf/10.1126/science.aaa7432}
  {https://www.science.org/doi/pdf/10.1126/science.aaa7432} \BibitemShut
  {NoStop}%
\bibitem [{\citenamefont {Sels}\ and\ \citenamefont
  {Polkovnikov}(2021)}]{Sels21}%
  \BibitemOpen
  \bibfield  {author} {\bibinfo {author} {\bibfnamefont {D.}~\bibnamefont
  {Sels}}\ and\ \bibinfo {author} {\bibfnamefont {A.}~\bibnamefont
  {Polkovnikov}},\ }\bibfield  {title} {\bibinfo {title} {Dynamical obstruction
  to localization in a disordered spin chain},\ }\href
  {https://doi.org/10.1103/PhysRevE.104.054105} {\bibfield  {journal} {\bibinfo
   {journal} {Phys. Rev. E}\ }\textbf {\bibinfo {volume} {104}},\ \bibinfo
  {pages} {054105} (\bibinfo {year} {2021})}\BibitemShut {NoStop}%
\bibitem [{\citenamefont {Kiefer-Emmanouilidis}\ \emph
  {et~al.}(2021)\citenamefont {Kiefer-Emmanouilidis}, \citenamefont {Unanyan},
  \citenamefont {Fleischhauer},\ and\ \citenamefont {Sirker}}]{Kiefer21}%
  \BibitemOpen
  \bibfield  {author} {\bibinfo {author} {\bibfnamefont {M.}~\bibnamefont
  {Kiefer-Emmanouilidis}}, \bibinfo {author} {\bibfnamefont {R.}~\bibnamefont
  {Unanyan}}, \bibinfo {author} {\bibfnamefont {M.}~\bibnamefont
  {Fleischhauer}},\ and\ \bibinfo {author} {\bibfnamefont {J.}~\bibnamefont
  {Sirker}},\ }\bibfield  {title} {\bibinfo {title} {Slow delocalization of
  particles in many-body localized phases},\ }\href
  {https://doi.org/10.1103/PhysRevB.103.024203} {\bibfield  {journal} {\bibinfo
   {journal} {Phys. Rev. B}\ }\textbf {\bibinfo {volume} {103}},\ \bibinfo
  {pages} {024203} (\bibinfo {year} {2021})}\BibitemShut {NoStop}%
\bibitem [{\citenamefont {Sels}\ and\ \citenamefont
  {Polkovnikov}(2023)}]{Sels23}%
  \BibitemOpen
  \bibfield  {author} {\bibinfo {author} {\bibfnamefont {D.}~\bibnamefont
  {Sels}}\ and\ \bibinfo {author} {\bibfnamefont {A.}~\bibnamefont
  {Polkovnikov}},\ }\bibfield  {title} {\bibinfo {title} {Thermalization of
  dilute impurities in one-dimensional spin chains},\ }\href
  {https://doi.org/10.1103/PhysRevX.13.011041} {\bibfield  {journal} {\bibinfo
  {journal} {Phys. Rev. X}\ }\textbf {\bibinfo {volume} {13}},\ \bibinfo
  {pages} {011041} (\bibinfo {year} {2023})}\BibitemShut {NoStop}%
\bibitem [{\citenamefont {Sierant}\ \emph {et~al.}(2025)\citenamefont
  {Sierant}, \citenamefont {Lewenstein}, \citenamefont {Scardicchio},
  \citenamefont {Vidmar},\ and\ \citenamefont {Zakrzewski}}]{Sierant25}%
  \BibitemOpen
  \bibfield  {author} {\bibinfo {author} {\bibfnamefont {P.}~\bibnamefont
  {Sierant}}, \bibinfo {author} {\bibfnamefont {M.}~\bibnamefont {Lewenstein}},
  \bibinfo {author} {\bibfnamefont {A.}~\bibnamefont {Scardicchio}}, \bibinfo
  {author} {\bibfnamefont {L.}~\bibnamefont {Vidmar}},\ and\ \bibinfo {author}
  {\bibfnamefont {J.}~\bibnamefont {Zakrzewski}},\ }\bibfield  {title}
  {\bibinfo {title} {Many-body localization in the age of classical
  computing*},\ }\href {https://doi.org/10.1088/1361-6633/ad9756} {\bibfield
  {journal} {\bibinfo  {journal} {Reports on Progress in Physics}\ }\textbf
  {\bibinfo {volume} {88}},\ \bibinfo {pages} {026502} (\bibinfo {year}
  {2025})}\BibitemShut {NoStop}%
\bibitem [{\citenamefont {Sierant}\ \emph {et~al.}(2020)\citenamefont
  {Sierant}, \citenamefont {Lewenstein},\ and\ \citenamefont
  {Zakrzewski}}]{Sierant20}%
  \BibitemOpen
  \bibfield  {author} {\bibinfo {author} {\bibfnamefont {P.}~\bibnamefont
  {Sierant}}, \bibinfo {author} {\bibfnamefont {M.}~\bibnamefont
  {Lewenstein}},\ and\ \bibinfo {author} {\bibfnamefont {J.}~\bibnamefont
  {Zakrzewski}},\ }\bibfield  {title} {\bibinfo {title} {Polynomially filtered
  exact diagonalization approach to many-body localization},\ }\href
  {https://doi.org/10.1103/PhysRevLett.125.156601} {\bibfield  {journal}
  {\bibinfo  {journal} {Phys. Rev. Lett.}\ }\textbf {\bibinfo {volume} {125}},\
  \bibinfo {pages} {156601} (\bibinfo {year} {2020})}\BibitemShut {NoStop}%
\bibitem [{\citenamefont {L\"uschen}\ \emph {et~al.}(2017)\citenamefont
  {L\"uschen}, \citenamefont {Bordia}, \citenamefont {Scherg}, \citenamefont
  {Alet}, \citenamefont {Altman}, \citenamefont {Schneider},\ and\
  \citenamefont {Bloch}}]{Luschen17}%
  \BibitemOpen
  \bibfield  {author} {\bibinfo {author} {\bibfnamefont {H.~P.}\ \bibnamefont
  {L\"uschen}}, \bibinfo {author} {\bibfnamefont {P.}~\bibnamefont {Bordia}},
  \bibinfo {author} {\bibfnamefont {S.}~\bibnamefont {Scherg}}, \bibinfo
  {author} {\bibfnamefont {F.}~\bibnamefont {Alet}}, \bibinfo {author}
  {\bibfnamefont {E.}~\bibnamefont {Altman}}, \bibinfo {author} {\bibfnamefont
  {U.}~\bibnamefont {Schneider}},\ and\ \bibinfo {author} {\bibfnamefont
  {I.}~\bibnamefont {Bloch}},\ }\bibfield  {title} {\bibinfo {title}
  {Observation of slow dynamics near the many-body localization transition in
  one-dimensional quasiperiodic systems},\ }\href
  {https://doi.org/10.1103/PhysRevLett.119.260401} {\bibfield  {journal}
  {\bibinfo  {journal} {Phys. Rev. Lett.}\ }\textbf {\bibinfo {volume} {119}},\
  \bibinfo {pages} {260401} (\bibinfo {year} {2017})}\BibitemShut {NoStop}%
\bibitem [{\citenamefont {Rispoli}\ \emph {et~al.}(2019)\citenamefont
  {Rispoli}, \citenamefont {Lukin}, \citenamefont {Schittko}, \citenamefont
  {Kim}, \citenamefont {Tai}, \citenamefont {Léonard},\ and\ \citenamefont
  {Greiner}}]{Rispoli19}%
  \BibitemOpen
  \bibfield  {author} {\bibinfo {author} {\bibfnamefont {M.}~\bibnamefont
  {Rispoli}}, \bibinfo {author} {\bibfnamefont {A.}~\bibnamefont {Lukin}},
  \bibinfo {author} {\bibfnamefont {R.}~\bibnamefont {Schittko}}, \bibinfo
  {author} {\bibfnamefont {S.-S.}\ \bibnamefont {Kim}}, \bibinfo {author}
  {\bibfnamefont {M.}~\bibnamefont {Tai}}, \bibinfo {author} {\bibfnamefont
  {J.}~\bibnamefont {Léonard}},\ and\ \bibinfo {author} {\bibfnamefont
  {M.}~\bibnamefont {Greiner}},\ }\bibfield  {title} {\bibinfo {title} {Quantum
  critical behaviour at the many-body localization transition},\ }\href
  {https://doi.org/10.1038/s41586-019-1527-2} {\bibfield  {journal} {\bibinfo
  {journal} {Nature}\ }\textbf {\bibinfo {volume} {573}},\ \bibinfo {pages}
  {385} (\bibinfo {year} {2019})}\BibitemShut {NoStop}%
\bibitem [{\citenamefont {L{\'e}onard}\ \emph {et~al.}(2023)\citenamefont
  {L{\'e}onard}, \citenamefont {Kim}, \citenamefont {Rispoli}, \citenamefont
  {Lukin}, \citenamefont {Schittko}, \citenamefont {Kwan}, \citenamefont
  {Demler}, \citenamefont {Sels},\ and\ \citenamefont {Greiner}}]{Leonard23}%
  \BibitemOpen
  \bibfield  {author} {\bibinfo {author} {\bibfnamefont {J.}~\bibnamefont
  {L{\'e}onard}}, \bibinfo {author} {\bibfnamefont {S.-S.}\ \bibnamefont
  {Kim}}, \bibinfo {author} {\bibfnamefont {M.}~\bibnamefont {Rispoli}},
  \bibinfo {author} {\bibfnamefont {A.}~\bibnamefont {Lukin}}, \bibinfo
  {author} {\bibfnamefont {R.}~\bibnamefont {Schittko}}, \bibinfo {author}
  {\bibfnamefont {J.}~\bibnamefont {Kwan}}, \bibinfo {author} {\bibfnamefont
  {E.}~\bibnamefont {Demler}}, \bibinfo {author} {\bibfnamefont
  {D.}~\bibnamefont {Sels}},\ and\ \bibinfo {author} {\bibfnamefont
  {M.}~\bibnamefont {Greiner}},\ }\bibfield  {title} {\bibinfo {title} {Probing
  the onset of quantum avalanches in a many-body localized system},\ }\href
  {https://doi.org/10.1038/s41567-022-01887-3} {\bibfield  {journal} {\bibinfo
  {journal} {Nature Physics}\ }\textbf {\bibinfo {volume} {19}},\ \bibinfo
  {pages} {481} (\bibinfo {year} {2023})}\BibitemShut {NoStop}%
\bibitem [{\citenamefont {Bera}\ \emph {et~al.}(2015)\citenamefont {Bera},
  \citenamefont {Schomerus}, \citenamefont {Heidrich-Meisner},\ and\
  \citenamefont {Bardarson}}]{Bera15}%
  \BibitemOpen
  \bibfield  {author} {\bibinfo {author} {\bibfnamefont {S.}~\bibnamefont
  {Bera}}, \bibinfo {author} {\bibfnamefont {H.}~\bibnamefont {Schomerus}},
  \bibinfo {author} {\bibfnamefont {F.}~\bibnamefont {Heidrich-Meisner}},\ and\
  \bibinfo {author} {\bibfnamefont {J.~H.}\ \bibnamefont {Bardarson}},\
  }\bibfield  {title} {\bibinfo {title} {Many-body localization characterized
  from a one-particle perspective},\ }\href
  {https://doi.org/10.1103/PhysRevLett.115.046603} {\bibfield  {journal}
  {\bibinfo  {journal} {Phys. Rev. Lett.}\ }\textbf {\bibinfo {volume} {115}},\
  \bibinfo {pages} {046603} (\bibinfo {year} {2015})}\BibitemShut {NoStop}%
\bibitem [{\citenamefont {Pal}\ and\ \citenamefont {Huse}(2010)}]{Pal10}%
  \BibitemOpen
  \bibfield  {author} {\bibinfo {author} {\bibfnamefont {A.}~\bibnamefont
  {Pal}}\ and\ \bibinfo {author} {\bibfnamefont {D.~A.}\ \bibnamefont {Huse}},\
  }\bibfield  {title} {\bibinfo {title} {Many-body localization phase
  transition},\ }\href {https://doi.org/10.1103/PhysRevB.82.174411} {\bibfield
  {journal} {\bibinfo  {journal} {Phys. Rev. B}\ }\textbf {\bibinfo {volume}
  {82}},\ \bibinfo {pages} {174411} (\bibinfo {year} {2010})}\BibitemShut
  {NoStop}%
\bibitem [{\citenamefont {Weisse}\ \emph {et~al.}(2025)\citenamefont {Weisse},
  \citenamefont {Gerstner},\ and\ \citenamefont {Sirker}}]{Weisse25}%
  \BibitemOpen
  \bibfield  {author} {\bibinfo {author} {\bibfnamefont {A.}~\bibnamefont
  {Weisse}}, \bibinfo {author} {\bibfnamefont {R.}~\bibnamefont {Gerstner}},\
  and\ \bibinfo {author} {\bibfnamefont {J.}~\bibnamefont {Sirker}},\
  }\bibfield  {title} {\bibinfo {title} {Operator growth in disordered spin
  chains: Indications for the absence of many-body localization},\ }\href
  {https://doi.org/10.1103/wgss-nt8t} {\bibfield  {journal} {\bibinfo
  {journal} {Phys. Rev. Res.}\ }\textbf {\bibinfo {volume} {7}},\ \bibinfo
  {pages} {033018} (\bibinfo {year} {2025})}\BibitemShut {NoStop}%
\bibitem [{\citenamefont {Serbyn}\ \emph
  {et~al.}(2013{\natexlab{b}})\citenamefont {Serbyn}, \citenamefont
  {Papi\ifmmode~\acute{c}\else \'{c}\fi{}},\ and\ \citenamefont
  {Abanin}}]{Serbyn13b}%
  \BibitemOpen
  \bibfield  {author} {\bibinfo {author} {\bibfnamefont {M.}~\bibnamefont
  {Serbyn}}, \bibinfo {author} {\bibfnamefont {Z.}~\bibnamefont
  {Papi\ifmmode~\acute{c}\else \'{c}\fi{}}},\ and\ \bibinfo {author}
  {\bibfnamefont {D.~A.}\ \bibnamefont {Abanin}},\ }\bibfield  {title}
  {\bibinfo {title} {Local conservation laws and the structure of the many-body
  localized states},\ }\href {https://doi.org/10.1103/PhysRevLett.111.127201}
  {\bibfield  {journal} {\bibinfo  {journal} {Phys. Rev. Lett.}\ }\textbf
  {\bibinfo {volume} {111}},\ \bibinfo {pages} {127201} (\bibinfo {year}
  {2013}{\natexlab{b}})}\BibitemShut {NoStop}%
\bibitem [{\citenamefont {Huse}\ \emph {et~al.}(2014)\citenamefont {Huse},
  \citenamefont {Nandkishore},\ and\ \citenamefont {Oganesyan}}]{Huse14}%
  \BibitemOpen
  \bibfield  {author} {\bibinfo {author} {\bibfnamefont {D.~A.}\ \bibnamefont
  {Huse}}, \bibinfo {author} {\bibfnamefont {R.}~\bibnamefont {Nandkishore}},\
  and\ \bibinfo {author} {\bibfnamefont {V.}~\bibnamefont {Oganesyan}},\
  }\bibfield  {title} {\bibinfo {title} {Phenomenology of fully
  many-body-localized systems},\ }\href
  {https://doi.org/10.1103/PhysRevB.90.174202} {\bibfield  {journal} {\bibinfo
  {journal} {Phys. Rev. B}\ }\textbf {\bibinfo {volume} {90}},\ \bibinfo
  {pages} {174202} (\bibinfo {year} {2014})}\BibitemShut {NoStop}%
\bibitem [{\citenamefont {Chandran}\ \emph {et~al.}(2015)\citenamefont
  {Chandran}, \citenamefont {Kim}, \citenamefont {Vidal},\ and\ \citenamefont
  {Abanin}}]{Chandran15}%
  \BibitemOpen
  \bibfield  {author} {\bibinfo {author} {\bibfnamefont {A.}~\bibnamefont
  {Chandran}}, \bibinfo {author} {\bibfnamefont {I.~H.}\ \bibnamefont {Kim}},
  \bibinfo {author} {\bibfnamefont {G.}~\bibnamefont {Vidal}},\ and\ \bibinfo
  {author} {\bibfnamefont {D.~A.}\ \bibnamefont {Abanin}},\ }\bibfield  {title}
  {\bibinfo {title} {Constructing local integrals of motion in the many-body
  localized phase},\ }\href {https://doi.org/10.1103/PhysRevB.91.085425}
  {\bibfield  {journal} {\bibinfo  {journal} {Phys. Rev. B}\ }\textbf {\bibinfo
  {volume} {91}},\ \bibinfo {pages} {085425} (\bibinfo {year}
  {2015})}\BibitemShut {NoStop}%
\bibitem [{\citenamefont {Rademaker}\ and\ \citenamefont
  {Ortu\~no}(2016)}]{Rademaker16}%
  \BibitemOpen
  \bibfield  {author} {\bibinfo {author} {\bibfnamefont {L.}~\bibnamefont
  {Rademaker}}\ and\ \bibinfo {author} {\bibfnamefont {M.}~\bibnamefont
  {Ortu\~no}},\ }\bibfield  {title} {\bibinfo {title} {Explicit local integrals
  of motion for the many-body localized state},\ }\href
  {https://doi.org/10.1103/PhysRevLett.116.010404} {\bibfield  {journal}
  {\bibinfo  {journal} {Phys. Rev. Lett.}\ }\textbf {\bibinfo {volume} {116}},\
  \bibinfo {pages} {010404} (\bibinfo {year} {2016})}\BibitemShut {NoStop}%
\bibitem [{\citenamefont {Thomson}\ and\ \citenamefont
  {Schir\'o}(2018)}]{Thomson18}%
  \BibitemOpen
  \bibfield  {author} {\bibinfo {author} {\bibfnamefont {S.~J.}\ \bibnamefont
  {Thomson}}\ and\ \bibinfo {author} {\bibfnamefont {M.}~\bibnamefont
  {Schir\'o}},\ }\bibfield  {title} {\bibinfo {title} {Time evolution of
  many-body localized systems with the flow equation approach},\ }\href
  {https://doi.org/10.1103/PhysRevB.97.060201} {\bibfield  {journal} {\bibinfo
  {journal} {Phys. Rev. B}\ }\textbf {\bibinfo {volume} {97}},\ \bibinfo
  {pages} {060201} (\bibinfo {year} {2018})}\BibitemShut {NoStop}%
\bibitem [{\citenamefont {Pekker}\ \emph {et~al.}(2017)\citenamefont {Pekker},
  \citenamefont {Clark}, \citenamefont {Oganesyan},\ and\ \citenamefont
  {Refael}}]{Pekker17}%
  \BibitemOpen
  \bibfield  {author} {\bibinfo {author} {\bibfnamefont {D.}~\bibnamefont
  {Pekker}}, \bibinfo {author} {\bibfnamefont {B.~K.}\ \bibnamefont {Clark}},
  \bibinfo {author} {\bibfnamefont {V.}~\bibnamefont {Oganesyan}},\ and\
  \bibinfo {author} {\bibfnamefont {G.}~\bibnamefont {Refael}},\ }\bibfield
  {title} {\bibinfo {title} {Fixed points of wegner-wilson flows and many-body
  localization},\ }\href {https://doi.org/10.1103/PhysRevLett.119.075701}
  {\bibfield  {journal} {\bibinfo  {journal} {Phys. Rev. Lett.}\ }\textbf
  {\bibinfo {volume} {119}},\ \bibinfo {pages} {075701} (\bibinfo {year}
  {2017})}\BibitemShut {NoStop}%
\bibitem [{\citenamefont {Resta}(1998)}]{Resta98}%
  \BibitemOpen
  \bibfield  {author} {\bibinfo {author} {\bibfnamefont {R.}~\bibnamefont
  {Resta}},\ }\bibfield  {title} {\bibinfo {title} {Quantum-mechanical position
  operator in extended systems},\ }\href
  {https://doi.org/10.1103/PhysRevLett.80.1800} {\bibfield  {journal} {\bibinfo
   {journal} {Phys. Rev. Lett.}\ }\textbf {\bibinfo {volume} {80}},\ \bibinfo
  {pages} {1800} (\bibinfo {year} {1998})}\BibitemShut {NoStop}%
\bibitem [{\citenamefont {Resta}\ and\ \citenamefont
  {Sorella}(1999)}]{Resta99}%
  \BibitemOpen
  \bibfield  {author} {\bibinfo {author} {\bibfnamefont {R.}~\bibnamefont
  {Resta}}\ and\ \bibinfo {author} {\bibfnamefont {S.}~\bibnamefont
  {Sorella}},\ }\bibfield  {title} {\bibinfo {title} {Electron localization in
  the insulating state},\ }\href {https://doi.org/10.1103/PhysRevLett.82.370}
  {\bibfield  {journal} {\bibinfo  {journal} {Phys. Rev. Lett.}\ }\textbf
  {\bibinfo {volume} {82}},\ \bibinfo {pages} {370} (\bibinfo {year}
  {1999})}\BibitemShut {NoStop}%
\bibitem [{\citenamefont {Souza}\ \emph {et~al.}(2000)\citenamefont {Souza},
  \citenamefont {Wilkens},\ and\ \citenamefont {Martin}}]{Souza00}%
  \BibitemOpen
  \bibfield  {author} {\bibinfo {author} {\bibfnamefont {I.}~\bibnamefont
  {Souza}}, \bibinfo {author} {\bibfnamefont {T.}~\bibnamefont {Wilkens}},\
  and\ \bibinfo {author} {\bibfnamefont {R.~M.}\ \bibnamefont {Martin}},\
  }\bibfield  {title} {\bibinfo {title} {Polarization and localization in
  insulators: Generating function approach},\ }\href
  {https://doi.org/10.1103/PhysRevB.62.1666} {\bibfield  {journal} {\bibinfo
  {journal} {Phys. Rev. B}\ }\textbf {\bibinfo {volume} {62}},\ \bibinfo
  {pages} {1666} (\bibinfo {year} {2000})}\BibitemShut {NoStop}%
\bibitem [{\citenamefont {Resta}(2002)}]{Resta02}%
  \BibitemOpen
  \bibfield  {author} {\bibinfo {author} {\bibfnamefont {R.}~\bibnamefont
  {Resta}},\ }\bibfield  {title} {\bibinfo {title} {Why are insulators
  insulating and metals conducting?},\ }\href
  {https://doi.org/10.1088/0953-8984/14/20/201} {\bibfield  {journal} {\bibinfo
   {journal} {Journal of Physics: Condensed Matter}\ }\textbf {\bibinfo
  {volume} {14}},\ \bibinfo {pages} {R625} (\bibinfo {year}
  {2002})}\BibitemShut {NoStop}%
\bibitem [{\citenamefont {Lu}\ and\ \citenamefont {Wang}(2010)}]{Lu10}%
  \BibitemOpen
  \bibfield  {author} {\bibinfo {author} {\bibfnamefont {X.-M.}\ \bibnamefont
  {Lu}}\ and\ \bibinfo {author} {\bibfnamefont {X.}~\bibnamefont {Wang}},\
  }\bibfield  {title} {\bibinfo {title} {Operator quantum geometric tensor and
  quantum phase transitions},\ }\href
  {https://doi.org/10.1209/0295-5075/91/30003} {\bibfield  {journal} {\bibinfo
  {journal} {Europhysics Letters}\ }\textbf {\bibinfo {volume} {91}},\ \bibinfo
  {pages} {30003} (\bibinfo {year} {2010})}\BibitemShut {NoStop}%
\bibitem [{\citenamefont {Valen\ifmmode \mbox{\c{c}}\else \c{c}\fi{}a
  Ferreira~de Arag\~ao}\ \emph {et~al.}(2019)\citenamefont {Valen\ifmmode
  \mbox{\c{c}}\else \c{c}\fi{}a Ferreira~de Arag\~ao}, \citenamefont {Moreno},
  \citenamefont {Battaglia}, \citenamefont {Bendazzoli}, \citenamefont
  {Evangelisti}, \citenamefont {Leininger}, \citenamefont {Suaud},\ and\
  \citenamefont {Berger}}]{Valenca19}%
  \BibitemOpen
  \bibfield  {author} {\bibinfo {author} {\bibfnamefont {E.}~\bibnamefont
  {Valen\ifmmode \mbox{\c{c}}\else \c{c}\fi{}a Ferreira~de Arag\~ao}}, \bibinfo
  {author} {\bibfnamefont {D.}~\bibnamefont {Moreno}}, \bibinfo {author}
  {\bibfnamefont {S.}~\bibnamefont {Battaglia}}, \bibinfo {author}
  {\bibfnamefont {G.~L.}\ \bibnamefont {Bendazzoli}}, \bibinfo {author}
  {\bibfnamefont {S.}~\bibnamefont {Evangelisti}}, \bibinfo {author}
  {\bibfnamefont {T.}~\bibnamefont {Leininger}}, \bibinfo {author}
  {\bibfnamefont {N.}~\bibnamefont {Suaud}},\ and\ \bibinfo {author}
  {\bibfnamefont {J.~A.}\ \bibnamefont {Berger}},\ }\bibfield  {title}
  {\bibinfo {title} {A simple position operator for periodic systems},\ }\href
  {https://doi.org/10.1103/PhysRevB.99.205144} {\bibfield  {journal} {\bibinfo
  {journal} {Phys. Rev. B}\ }\textbf {\bibinfo {volume} {99}},\ \bibinfo
  {pages} {205144} (\bibinfo {year} {2019})}\BibitemShut {NoStop}%
\bibitem [{\citenamefont {Het\'enyi}\ and\ \citenamefont
  {D\'ora}(2019)}]{Hetenyi19}%
  \BibitemOpen
  \bibfield  {author} {\bibinfo {author} {\bibfnamefont {B.}~\bibnamefont
  {Het\'enyi}}\ and\ \bibinfo {author} {\bibfnamefont {B.}~\bibnamefont
  {D\'ora}},\ }\bibfield  {title} {\bibinfo {title} {Quantum phase transitions
  from analysis of the polarization amplitude},\ }\href
  {https://doi.org/10.1103/PhysRevB.99.085126} {\bibfield  {journal} {\bibinfo
  {journal} {Phys. Rev. B}\ }\textbf {\bibinfo {volume} {99}},\ \bibinfo
  {pages} {085126} (\bibinfo {year} {2019})}\BibitemShut {NoStop}%
\bibitem [{\citenamefont {Het\'enyi}\ and\ \citenamefont
  {Cengiz}(2022)}]{Hetenyi22}%
  \BibitemOpen
  \bibfield  {author} {\bibinfo {author} {\bibfnamefont {B.}~\bibnamefont
  {Het\'enyi}}\ and\ \bibinfo {author} {\bibfnamefont {S.~m.~c.}\ \bibnamefont
  {Cengiz}},\ }\bibfield  {title} {\bibinfo {title} {Geometric cumulants
  associated with adiabatic cycles crossing degeneracy points: Application to
  finite size scaling of metal-insulator transitions in crystalline electronic
  systems},\ }\href {https://doi.org/10.1103/PhysRevB.106.195151} {\bibfield
  {journal} {\bibinfo  {journal} {Phys. Rev. B}\ }\textbf {\bibinfo {volume}
  {106}},\ \bibinfo {pages} {195151} (\bibinfo {year} {2022})}\BibitemShut
  {NoStop}%
\bibitem [{\citenamefont {Resta}(2005)}]{Resta05}%
  \BibitemOpen
  \bibfield  {author} {\bibinfo {author} {\bibfnamefont {R.}~\bibnamefont
  {Resta}},\ }\bibfield  {title} {\bibinfo {title} {Electron localization in
  the quantum hall regime},\ }\href
  {https://doi.org/10.1103/PhysRevLett.95.196805} {\bibfield  {journal}
  {\bibinfo  {journal} {Phys. Rev. Lett.}\ }\textbf {\bibinfo {volume} {95}},\
  \bibinfo {pages} {196805} (\bibinfo {year} {2005})}\BibitemShut {NoStop}%
\bibitem [{\citenamefont {Wilkens}\ and\ \citenamefont
  {Martin}(2001)}]{Wilkens01}%
  \BibitemOpen
  \bibfield  {author} {\bibinfo {author} {\bibfnamefont {T.}~\bibnamefont
  {Wilkens}}\ and\ \bibinfo {author} {\bibfnamefont {R.~M.}\ \bibnamefont
  {Martin}},\ }\bibfield  {title} {\bibinfo {title} {Quantum monte carlo study
  of the one-dimensional ionic hubbard model},\ }\href
  {https://doi.org/10.1103/PhysRevB.63.235108} {\bibfield  {journal} {\bibinfo
  {journal} {Phys. Rev. B}\ }\textbf {\bibinfo {volume} {63}},\ \bibinfo
  {pages} {235108} (\bibinfo {year} {2001})}\BibitemShut {NoStop}%
\bibitem [{\citenamefont {Thonhauser}\ and\ \citenamefont
  {Vanderbilt}(2006)}]{Thonhauser06}%
  \BibitemOpen
  \bibfield  {author} {\bibinfo {author} {\bibfnamefont {T.}~\bibnamefont
  {Thonhauser}}\ and\ \bibinfo {author} {\bibfnamefont {D.}~\bibnamefont
  {Vanderbilt}},\ }\bibfield  {title} {\bibinfo {title}
  {Insulator/chern-insulator transition in the haldane model},\ }\href
  {https://doi.org/10.1103/PhysRevB.74.235111} {\bibfield  {journal} {\bibinfo
  {journal} {Phys. Rev. B}\ }\textbf {\bibinfo {volume} {74}},\ \bibinfo
  {pages} {235111} (\bibinfo {year} {2006})}\BibitemShut {NoStop}%
\bibitem [{\citenamefont {Marsal}\ \emph {et~al.}(2025)\citenamefont {Marsal},
  \citenamefont {Holmvall},\ and\ \citenamefont {Black-Schaffer}}]{Marsal25}%
  \BibitemOpen
  \bibfield  {author} {\bibinfo {author} {\bibfnamefont {Q.}~\bibnamefont
  {Marsal}}, \bibinfo {author} {\bibfnamefont {P.}~\bibnamefont {Holmvall}},\
  and\ \bibinfo {author} {\bibfnamefont {A.~M.}\ \bibnamefont
  {Black-Schaffer}},\ }\href {https://arxiv.org/abs/2506.15575} {\bibinfo
  {title} {Quantum metric and localization in a quasicrystal}} (\bibinfo {year}
  {2025}),\ \Eprint {https://arxiv.org/abs/2506.15575} {arXiv:2506.15575
  [cond-mat.mes-hall]} \BibitemShut {NoStop}%
\bibitem [{\citenamefont {Filippone}\ \emph {et~al.}(2016)\citenamefont
  {Filippone}, \citenamefont {Brouwer}, \citenamefont {Eisert},\ and\
  \citenamefont {von Oppen}}]{Filippone16}%
  \BibitemOpen
  \bibfield  {author} {\bibinfo {author} {\bibfnamefont {M.}~\bibnamefont
  {Filippone}}, \bibinfo {author} {\bibfnamefont {P.~W.}\ \bibnamefont
  {Brouwer}}, \bibinfo {author} {\bibfnamefont {J.}~\bibnamefont {Eisert}},\
  and\ \bibinfo {author} {\bibfnamefont {F.}~\bibnamefont {von Oppen}},\
  }\bibfield  {title} {\bibinfo {title} {Drude weight fluctuations in many-body
  localized systems},\ }\href {https://doi.org/10.1103/PhysRevB.94.201112}
  {\bibfield  {journal} {\bibinfo  {journal} {Phys. Rev. B}\ }\textbf {\bibinfo
  {volume} {94}},\ \bibinfo {pages} {201112} (\bibinfo {year}
  {2016})}\BibitemShut {NoStop}%
\bibitem [{\citenamefont {Pouranvari}\ and\ \citenamefont
  {Liou}(2021)}]{Pouranvari21}%
  \BibitemOpen
  \bibfield  {author} {\bibinfo {author} {\bibfnamefont {M.}~\bibnamefont
  {Pouranvari}}\ and\ \bibinfo {author} {\bibfnamefont {S.-F.}\ \bibnamefont
  {Liou}},\ }\bibfield  {title} {\bibinfo {title} {Characterizing many-body
  localization via state sensitivity to boundary conditions},\ }\href
  {https://doi.org/10.1103/PhysRevB.103.035136} {\bibfield  {journal} {\bibinfo
   {journal} {Phys. Rev. B}\ }\textbf {\bibinfo {volume} {103}},\ \bibinfo
  {pages} {035136} (\bibinfo {year} {2021})}\BibitemShut {NoStop}%
\bibitem [{\citenamefont {Hamazaki}\ \emph {et~al.}(2019)\citenamefont
  {Hamazaki}, \citenamefont {Kawabata},\ and\ \citenamefont
  {Ueda}}]{Hamazaki19}%
  \BibitemOpen
  \bibfield  {author} {\bibinfo {author} {\bibfnamefont {R.}~\bibnamefont
  {Hamazaki}}, \bibinfo {author} {\bibfnamefont {K.}~\bibnamefont {Kawabata}},\
  and\ \bibinfo {author} {\bibfnamefont {M.}~\bibnamefont {Ueda}},\ }\bibfield
  {title} {\bibinfo {title} {Non-hermitian many-body localization},\ }\href
  {https://doi.org/10.1103/PhysRevLett.123.090603} {\bibfield  {journal}
  {\bibinfo  {journal} {Phys. Rev. Lett.}\ }\textbf {\bibinfo {volume} {123}},\
  \bibinfo {pages} {090603} (\bibinfo {year} {2019})}\BibitemShut {NoStop}%
\bibitem [{\citenamefont {Heu\ss{}en}\ \emph {et~al.}(2021)\citenamefont
  {Heu\ss{}en}, \citenamefont {White},\ and\ \citenamefont
  {Refael}}]{Heuben21}%
  \BibitemOpen
  \bibfield  {author} {\bibinfo {author} {\bibfnamefont {S.}~\bibnamefont
  {Heu\ss{}en}}, \bibinfo {author} {\bibfnamefont {C.~D.}\ \bibnamefont
  {White}},\ and\ \bibinfo {author} {\bibfnamefont {G.}~\bibnamefont
  {Refael}},\ }\bibfield  {title} {\bibinfo {title} {Extracting many-body
  localization lengths with an imaginary vector potential},\ }\href
  {https://doi.org/10.1103/PhysRevB.103.064201} {\bibfield  {journal} {\bibinfo
   {journal} {Phys. Rev. B}\ }\textbf {\bibinfo {volume} {103}},\ \bibinfo
  {pages} {064201} (\bibinfo {year} {2021})}\BibitemShut {NoStop}%
\bibitem [{\citenamefont {O'Brien}\ and\ \citenamefont
  {Refael}(2023)}]{Obrien23}%
  \BibitemOpen
  \bibfield  {author} {\bibinfo {author} {\bibfnamefont {L.}~\bibnamefont
  {O'Brien}}\ and\ \bibinfo {author} {\bibfnamefont {G.}~\bibnamefont
  {Refael}},\ }\bibfield  {title} {\bibinfo {title} {Probing localization
  properties of many-body hamiltonians via an imaginary vector potential},\
  }\href {https://doi.org/10.1103/PhysRevB.108.184207} {\bibfield  {journal}
  {\bibinfo  {journal} {Phys. Rev. B}\ }\textbf {\bibinfo {volume} {108}},\
  \bibinfo {pages} {184207} (\bibinfo {year} {2023})}\BibitemShut {NoStop}%
\bibitem [{\citenamefont {Sierant}\ \emph {et~al.}(2019)\citenamefont
  {Sierant}, \citenamefont {Maksymov}, \citenamefont {Ku\ifmmode~\acute{s}\else
  \'{s}\fi{}},\ and\ \citenamefont {Zakrzewski}}]{Sierant19}%
  \BibitemOpen
  \bibfield  {author} {\bibinfo {author} {\bibfnamefont {P.}~\bibnamefont
  {Sierant}}, \bibinfo {author} {\bibfnamefont {A.}~\bibnamefont {Maksymov}},
  \bibinfo {author} {\bibfnamefont {M.}~\bibnamefont {Ku\ifmmode~\acute{s}\else
  \'{s}\fi{}}},\ and\ \bibinfo {author} {\bibfnamefont {J.}~\bibnamefont
  {Zakrzewski}},\ }\bibfield  {title} {\bibinfo {title} {Fidelity
  susceptibility in gaussian random ensembles},\ }\href
  {https://doi.org/10.1103/PhysRevE.99.050102} {\bibfield  {journal} {\bibinfo
  {journal} {Phys. Rev. E}\ }\textbf {\bibinfo {volume} {99}},\ \bibinfo
  {pages} {050102} (\bibinfo {year} {2019})}\BibitemShut {NoStop}%
\bibitem [{\citenamefont {Maksymov}\ \emph {et~al.}(2019)\citenamefont
  {Maksymov}, \citenamefont {Sierant},\ and\ \citenamefont
  {Zakrzewski}}]{Maksymov19}%
  \BibitemOpen
  \bibfield  {author} {\bibinfo {author} {\bibfnamefont {A.}~\bibnamefont
  {Maksymov}}, \bibinfo {author} {\bibfnamefont {P.}~\bibnamefont {Sierant}},\
  and\ \bibinfo {author} {\bibfnamefont {J.}~\bibnamefont {Zakrzewski}},\
  }\bibfield  {title} {\bibinfo {title} {Energy level dynamics across the
  many-body localization transition},\ }\href
  {https://doi.org/10.1103/PhysRevB.99.224202} {\bibfield  {journal} {\bibinfo
  {journal} {Phys. Rev. B}\ }\textbf {\bibinfo {volume} {99}},\ \bibinfo
  {pages} {224202} (\bibinfo {year} {2019})}\BibitemShut {NoStop}%
\bibitem [{\citenamefont {Resta}(2011)}]{Resta11}%
  \BibitemOpen
  \bibfield  {author} {\bibinfo {author} {\bibfnamefont {R.}~\bibnamefont
  {Resta}},\ }\bibfield  {title} {\bibinfo {title} {The insulating state of
  matter: a geometrical theory},\ }\href
  {https://link.springer.com/article/10.1140/epjb/e2010-10874-4} {\bibfield
  {journal} {\bibinfo  {journal} {The European Physical Journal B}\ }\textbf
  {\bibinfo {volume} {79}},\ \bibinfo {pages} {121} (\bibinfo {year}
  {2011})}\BibitemShut {NoStop}%
\bibitem [{\citenamefont {Resta}(2020)}]{Resta20}%
  \BibitemOpen
  \bibfield  {author} {\bibinfo {author} {\bibfnamefont {R.}~\bibnamefont
  {Resta}},\ }\href {https://arxiv.org/abs/2006.15567} {\bibinfo {title}
  {Geometry and topology in many-body physics}} (\bibinfo {year} {2020}),\
  \Eprint {https://arxiv.org/abs/2006.15567} {arXiv:2006.15567
  [cond-mat.str-el]} \BibitemShut {NoStop}%
\bibitem [{\citenamefont {Niu}\ \emph {et~al.}(1985)\citenamefont {Niu},
  \citenamefont {Thouless},\ and\ \citenamefont {Wu}}]{Niu85}%
  \BibitemOpen
  \bibfield  {author} {\bibinfo {author} {\bibfnamefont {Q.}~\bibnamefont
  {Niu}}, \bibinfo {author} {\bibfnamefont {D.~J.}\ \bibnamefont {Thouless}},\
  and\ \bibinfo {author} {\bibfnamefont {Y.-S.}\ \bibnamefont {Wu}},\
  }\bibfield  {title} {\bibinfo {title} {Quantized hall conductance as a
  topological invariant},\ }\href {https://doi.org/10.1103/PhysRevB.31.3372}
  {\bibfield  {journal} {\bibinfo  {journal} {Phys. Rev. B}\ }\textbf {\bibinfo
  {volume} {31}},\ \bibinfo {pages} {3372} (\bibinfo {year}
  {1985})}\BibitemShut {NoStop}%
\bibitem [{\citenamefont {Banerjee}(1996)}]{Banerjee96}%
  \BibitemOpen
  \bibfield  {author} {\bibinfo {author} {\bibfnamefont {D.}~\bibnamefont
  {Banerjee}},\ }\bibfield  {title} {\bibinfo {title} {Topological aspects of
  the berry phase},\ }\href
  {https://doi.org/https://doi.org/10.1002/prop.2190440403} {\bibfield
  {journal} {\bibinfo  {journal} {Fortschritte der Physik/Progress of Physics}\
  }\textbf {\bibinfo {volume} {44}},\ \bibinfo {pages} {323} (\bibinfo {year}
  {1996})},\ \Eprint
  {https://arxiv.org/abs/https://onlinelibrary.wiley.com/doi/pdf/10.1002/prop.2190440403}
  {https://onlinelibrary.wiley.com/doi/pdf/10.1002/prop.2190440403}
  \BibitemShut {NoStop}%
\bibitem [{\citenamefont {Campos~Venuti}\ and\ \citenamefont
  {Zanardi}(2007)}]{Venuti07}%
  \BibitemOpen
  \bibfield  {author} {\bibinfo {author} {\bibfnamefont {L.}~\bibnamefont
  {Campos~Venuti}}\ and\ \bibinfo {author} {\bibfnamefont {P.}~\bibnamefont
  {Zanardi}},\ }\bibfield  {title} {\bibinfo {title} {Quantum critical scaling
  of the geometric tensors},\ }\href
  {https://doi.org/10.1103/PhysRevLett.99.095701} {\bibfield  {journal}
  {\bibinfo  {journal} {Phys. Rev. Lett.}\ }\textbf {\bibinfo {volume} {99}},\
  \bibinfo {pages} {095701} (\bibinfo {year} {2007})}\BibitemShut {NoStop}%
\bibitem [{\citenamefont {Watanabe}(2018)}]{Watanabe18}%
  \BibitemOpen
  \bibfield  {author} {\bibinfo {author} {\bibfnamefont {H.}~\bibnamefont
  {Watanabe}},\ }\bibfield  {title} {\bibinfo {title} {Insensitivity of bulk
  properties to the twisted boundary condition},\ }\href
  {https://doi.org/10.1103/PhysRevB.98.155137} {\bibfield  {journal} {\bibinfo
  {journal} {Phys. Rev. B}\ }\textbf {\bibinfo {volume} {98}},\ \bibinfo
  {pages} {155137} (\bibinfo {year} {2018})}\BibitemShut {NoStop}%
\bibitem [{\citenamefont {Ozawa}\ and\ \citenamefont
  {Goldman}(2019)}]{Ozawa19}%
  \BibitemOpen
  \bibfield  {author} {\bibinfo {author} {\bibfnamefont {T.}~\bibnamefont
  {Ozawa}}\ and\ \bibinfo {author} {\bibfnamefont {N.}~\bibnamefont
  {Goldman}},\ }\bibfield  {title} {\bibinfo {title} {Probing localization and
  quantum geometry by spectroscopy},\ }\href
  {https://doi.org/10.1103/PhysRevResearch.1.032019} {\bibfield  {journal}
  {\bibinfo  {journal} {Phys. Rev. Res.}\ }\textbf {\bibinfo {volume} {1}},\
  \bibinfo {pages} {032019} (\bibinfo {year} {2019})}\BibitemShut {NoStop}%
\bibitem [{\citenamefont {Ding}\ \emph {et~al.}(2024)\citenamefont {Ding},
  \citenamefont {Zhang}, \citenamefont {Liu}, \citenamefont {Wang},
  \citenamefont {Zhang},\ and\ \citenamefont {Zhu}}]{Ding24}%
  \BibitemOpen
  \bibfield  {author} {\bibinfo {author} {\bibfnamefont {H.-T.}\ \bibnamefont
  {Ding}}, \bibinfo {author} {\bibfnamefont {C.-X.}\ \bibnamefont {Zhang}},
  \bibinfo {author} {\bibfnamefont {J.-X.}\ \bibnamefont {Liu}}, \bibinfo
  {author} {\bibfnamefont {J.-T.}\ \bibnamefont {Wang}}, \bibinfo {author}
  {\bibfnamefont {D.-W.}\ \bibnamefont {Zhang}},\ and\ \bibinfo {author}
  {\bibfnamefont {S.-L.}\ \bibnamefont {Zhu}},\ }\bibfield  {title} {\bibinfo
  {title} {Non-abelian quantum geometric tensor in degenerate topological
  semimetals},\ }\href {https://doi.org/10.1103/PhysRevA.109.043305} {\bibfield
   {journal} {\bibinfo  {journal} {Phys. Rev. A}\ }\textbf {\bibinfo {volume}
  {109}},\ \bibinfo {pages} {043305} (\bibinfo {year} {2024})}\BibitemShut
  {NoStop}%
\bibitem [{\citenamefont {Ozawa}\ and\ \citenamefont {Mera}(2021)}]{Ozawa21}%
  \BibitemOpen
  \bibfield  {author} {\bibinfo {author} {\bibfnamefont {T.}~\bibnamefont
  {Ozawa}}\ and\ \bibinfo {author} {\bibfnamefont {B.}~\bibnamefont {Mera}},\
  }\bibfield  {title} {\bibinfo {title} {Relations between topology and the
  quantum metric for chern insulators},\ }\href
  {https://doi.org/10.1103/PhysRevB.104.045103} {\bibfield  {journal} {\bibinfo
   {journal} {Phys. Rev. B}\ }\textbf {\bibinfo {volume} {104}},\ \bibinfo
  {pages} {045103} (\bibinfo {year} {2021})}\BibitemShut {NoStop}%
\bibitem [{\citenamefont {Kondov}\ \emph {et~al.}(2015)\citenamefont {Kondov},
  \citenamefont {McGehee}, \citenamefont {Xu},\ and\ \citenamefont
  {DeMarco}}]{Kondov15}%
  \BibitemOpen
  \bibfield  {author} {\bibinfo {author} {\bibfnamefont {S.~S.}\ \bibnamefont
  {Kondov}}, \bibinfo {author} {\bibfnamefont {W.~R.}\ \bibnamefont {McGehee}},
  \bibinfo {author} {\bibfnamefont {W.}~\bibnamefont {Xu}},\ and\ \bibinfo
  {author} {\bibfnamefont {B.}~\bibnamefont {DeMarco}},\ }\bibfield  {title}
  {\bibinfo {title} {Disorder-induced localization in a strongly correlated
  atomic hubbard gas},\ }\href {https://doi.org/10.1103/PhysRevLett.114.083002}
  {\bibfield  {journal} {\bibinfo  {journal} {Phys. Rev. Lett.}\ }\textbf
  {\bibinfo {volume} {114}},\ \bibinfo {pages} {083002} (\bibinfo {year}
  {2015})}\BibitemShut {NoStop}%
\bibitem [{\citenamefont {Serbyn}\ \emph {et~al.}(2015)\citenamefont {Serbyn},
  \citenamefont {Papi\ifmmode~\acute{c}\else \'{c}\fi{}},\ and\ \citenamefont
  {Abanin}}]{Serbyn15}%
  \BibitemOpen
  \bibfield  {author} {\bibinfo {author} {\bibfnamefont {M.}~\bibnamefont
  {Serbyn}}, \bibinfo {author} {\bibfnamefont {Z.}~\bibnamefont
  {Papi\ifmmode~\acute{c}\else \'{c}\fi{}}},\ and\ \bibinfo {author}
  {\bibfnamefont {D.~A.}\ \bibnamefont {Abanin}},\ }\bibfield  {title}
  {\bibinfo {title} {Criterion for many-body localization-delocalization phase
  transition},\ }\href {https://doi.org/10.1103/PhysRevX.5.041047} {\bibfield
  {journal} {\bibinfo  {journal} {Phys. Rev. X}\ }\textbf {\bibinfo {volume}
  {5}},\ \bibinfo {pages} {041047} (\bibinfo {year} {2015})}\BibitemShut
  {NoStop}%
\bibitem [{\citenamefont {Mondaini}\ and\ \citenamefont
  {Rigol}(2015)}]{Mondaini15}%
  \BibitemOpen
  \bibfield  {author} {\bibinfo {author} {\bibfnamefont {R.}~\bibnamefont
  {Mondaini}}\ and\ \bibinfo {author} {\bibfnamefont {M.}~\bibnamefont
  {Rigol}},\ }\bibfield  {title} {\bibinfo {title} {Many-body localization and
  thermalization in disordered hubbard chains},\ }\href
  {https://doi.org/10.1103/PhysRevA.92.041601} {\bibfield  {journal} {\bibinfo
  {journal} {Phys. Rev. A}\ }\textbf {\bibinfo {volume} {92}},\ \bibinfo
  {pages} {041601} (\bibinfo {year} {2015})}\BibitemShut {NoStop}%
\bibitem [{\citenamefont {Chen}\ \emph {et~al.}(2023)\citenamefont {Chen},
  \citenamefont {Chen},\ and\ \citenamefont {Wang}}]{Chen23}%
  \BibitemOpen
  \bibfield  {author} {\bibinfo {author} {\bibfnamefont {J.}~\bibnamefont
  {Chen}}, \bibinfo {author} {\bibfnamefont {C.}~\bibnamefont {Chen}},\ and\
  \bibinfo {author} {\bibfnamefont {X.}~\bibnamefont {Wang}},\ }\href@noop {}
  {\bibinfo {title} {Many-body localization transition in the disordered
  bose-hubbard chain}} (\bibinfo {year} {2023}),\ \Eprint
  {https://arxiv.org/abs/2104.08582} {arXiv:2104.08582 [cond-mat.dis-nn]}
  \BibitemShut {NoStop}%
\bibitem [{\citenamefont {Wilkens}\ \emph {et~al.}(2023)\citenamefont
  {Wilkens}, \citenamefont {Lawrence}, \citenamefont {Walsh}, \citenamefont
  {Morita}, \citenamefont {Simpson}, \citenamefont {Ritter}, \citenamefont
  {Stenning}, \citenamefont {Arevalo-Lopez},\ and\ \citenamefont
  {Mclaughlin}}]{Wilkens23}%
  \BibitemOpen
  \bibfield  {author} {\bibinfo {author} {\bibfnamefont {E.~J.}\ \bibnamefont
  {Wilkens}}, \bibinfo {author} {\bibfnamefont {G.}~\bibnamefont {Lawrence}},
  \bibinfo {author} {\bibfnamefont {A.}~\bibnamefont {Walsh}}, \bibinfo
  {author} {\bibfnamefont {K.}~\bibnamefont {Morita}}, \bibinfo {author}
  {\bibfnamefont {S.}~\bibnamefont {Simpson}}, \bibinfo {author} {\bibfnamefont
  {C.}~\bibnamefont {Ritter}}, \bibinfo {author} {\bibfnamefont {G.~B.~G.}\
  \bibnamefont {Stenning}}, \bibinfo {author} {\bibfnamefont {A.~M.}\
  \bibnamefont {Arevalo-Lopez}},\ and\ \bibinfo {author} {\bibfnamefont
  {A.~C.}\ \bibnamefont {Mclaughlin}},\ }\bibfield  {title} {\bibinfo {title}
  {Observation of an exotic insulator to insulator transition upon electron
  doping the mott insulator cemnaso},\ }\href
  {https://doi.org/10.1038/s41467-023-42858-3} {\bibfield  {journal} {\bibinfo
  {journal} {Nat. Comm.}\ }\textbf {\bibinfo {volume} {14}},\ \bibinfo {pages}
  {7037} (\bibinfo {year} {2023})}\BibitemShut {NoStop}%
\bibitem [{\citenamefont {Asteria}\ \emph {et~al.}(2019)\citenamefont
  {Asteria}, \citenamefont {Tran}, \citenamefont {Ozawa}, \citenamefont
  {Tarnowski}, \citenamefont {Rem}, \citenamefont {Fl{\"a}schner},
  \citenamefont {Sengstock}, \citenamefont {Goldman},\ and\ \citenamefont
  {Weitenberg}}]{Asteria2019}%
  \BibitemOpen
  \bibfield  {author} {\bibinfo {author} {\bibfnamefont {L.}~\bibnamefont
  {Asteria}}, \bibinfo {author} {\bibfnamefont {D.~T.}\ \bibnamefont {Tran}},
  \bibinfo {author} {\bibfnamefont {T.}~\bibnamefont {Ozawa}}, \bibinfo
  {author} {\bibfnamefont {M.}~\bibnamefont {Tarnowski}}, \bibinfo {author}
  {\bibfnamefont {B.~S.}\ \bibnamefont {Rem}}, \bibinfo {author} {\bibfnamefont
  {N.}~\bibnamefont {Fl{\"a}schner}}, \bibinfo {author} {\bibfnamefont
  {K.}~\bibnamefont {Sengstock}}, \bibinfo {author} {\bibfnamefont
  {N.}~\bibnamefont {Goldman}},\ and\ \bibinfo {author} {\bibfnamefont
  {C.}~\bibnamefont {Weitenberg}},\ }\bibfield  {title} {\bibinfo {title}
  {Measuring quantized circular dichroism in ultracold topological matter},\
  }\href {https://www.nature.com/articles/s41567-019-0417-8} {\bibfield
  {journal} {\bibinfo  {journal} {Nature physics}\ }\textbf {\bibinfo {volume}
  {15}},\ \bibinfo {pages} {449} (\bibinfo {year} {2019})}\BibitemShut
  {NoStop}%
\bibitem [{\citenamefont {Onishi}\ and\ \citenamefont {Fu}(2025)}]{Onishi2025}%
  \BibitemOpen
  \bibfield  {author} {\bibinfo {author} {\bibfnamefont {Y.}~\bibnamefont
  {Onishi}}\ and\ \bibinfo {author} {\bibfnamefont {L.}~\bibnamefont {Fu}},\
  }\bibfield  {title} {\bibinfo {title} {Quantum weight: A fundamental property
  of quantum many-body systems},\ }\href
  {https://doi.org/10.1103/PhysRevResearch.7.023158} {\bibfield  {journal}
  {\bibinfo  {journal} {Phys. Rev. Res.}\ }\textbf {\bibinfo {volume} {7}},\
  \bibinfo {pages} {023158} (\bibinfo {year} {2025})}\BibitemShut {NoStop}%
\bibitem [{\citenamefont {Ba\l{}ut}\ \emph {et~al.}(2025)\citenamefont
  {Ba\l{}ut}, \citenamefont {Bradlyn},\ and\ \citenamefont
  {Abbamonte}}]{Balut2025}%
  \BibitemOpen
  \bibfield  {author} {\bibinfo {author} {\bibfnamefont {D.}~\bibnamefont
  {Ba\l{}ut}}, \bibinfo {author} {\bibfnamefont {B.}~\bibnamefont {Bradlyn}},\
  and\ \bibinfo {author} {\bibfnamefont {P.}~\bibnamefont {Abbamonte}},\
  }\bibfield  {title} {\bibinfo {title} {Quantum entanglement and quantum
  geometry measured with inelastic x-ray scattering},\ }\href
  {https://doi.org/10.1103/PhysRevB.111.125161} {\bibfield  {journal} {\bibinfo
   {journal} {Phys. Rev. B}\ }\textbf {\bibinfo {volume} {111}},\ \bibinfo
  {pages} {125161} (\bibinfo {year} {2025})}\BibitemShut {NoStop}%
\bibitem [{\citenamefont {Liu}\ \emph {et~al.}(2023)\citenamefont {Liu},
  \citenamefont {Zhang}, \citenamefont {Hsieh}, \citenamefont {Zhang},\ and\
  \citenamefont {Yao}}]{Liu23}%
  \BibitemOpen
  \bibfield  {author} {\bibinfo {author} {\bibfnamefont {S.}~\bibnamefont
  {Liu}}, \bibinfo {author} {\bibfnamefont {S.-X.}\ \bibnamefont {Zhang}},
  \bibinfo {author} {\bibfnamefont {C.-Y.}\ \bibnamefont {Hsieh}}, \bibinfo
  {author} {\bibfnamefont {S.}~\bibnamefont {Zhang}},\ and\ \bibinfo {author}
  {\bibfnamefont {H.}~\bibnamefont {Yao}},\ }\bibfield  {title} {\bibinfo
  {title} {Probing many-body localization by excited-state variational quantum
  eigensolver},\ }\href {https://doi.org/10.1103/PhysRevB.107.024204}
  {\bibfield  {journal} {\bibinfo  {journal} {Phys. Rev. B}\ }\textbf {\bibinfo
  {volume} {107}},\ \bibinfo {pages} {024204} (\bibinfo {year}
  {2023})}\BibitemShut {NoStop}%
\bibitem [{\citenamefont {Guo}\ and\ \citenamefont {Li}(2018)}]{Guo18}%
  \BibitemOpen
  \bibfield  {author} {\bibinfo {author} {\bibfnamefont {X.}~\bibnamefont
  {Guo}}\ and\ \bibinfo {author} {\bibfnamefont {X.}~\bibnamefont {Li}},\
  }\bibfield  {title} {\bibinfo {title} {Detecting many-body-localization
  lengths with cold atoms},\ }\href
  {https://doi.org/10.1103/PhysRevA.97.033622} {\bibfield  {journal} {\bibinfo
  {journal} {Phys. Rev. A}\ }\textbf {\bibinfo {volume} {97}},\ \bibinfo
  {pages} {033622} (\bibinfo {year} {2018})}\BibitemShut {NoStop}%
\bibitem [{\citenamefont {Romeral}\ \emph {et~al.}(2024)\citenamefont
  {Romeral}, \citenamefont {Cummings},\ and\ \citenamefont
  {Roche}}]{Romeral24}%
  \BibitemOpen
  \bibfield  {author} {\bibinfo {author} {\bibfnamefont {J.~M.}\ \bibnamefont
  {Romeral}}, \bibinfo {author} {\bibfnamefont {A.~W.}\ \bibnamefont
  {Cummings}},\ and\ \bibinfo {author} {\bibfnamefont {S.}~\bibnamefont
  {Roche}},\ }\href {https://arxiv.org/abs/2406.12677} {\bibinfo {title}
  {Scaling of the quantum geometry metrics in disordered topological phases}}
  (\bibinfo {year} {2024}),\ \Eprint {https://arxiv.org/abs/2406.12677}
  {arXiv:2406.12677 [cond-mat.dis-nn]} \BibitemShut {NoStop}%
\bibitem [{\citenamefont {Dumitrescu}\ \emph {et~al.}(2019)\citenamefont
  {Dumitrescu}, \citenamefont {Goremykina}, \citenamefont {Parameswaran},
  \citenamefont {Serbyn},\ and\ \citenamefont {Vasseur}}]{Dumitrescu19}%
  \BibitemOpen
  \bibfield  {author} {\bibinfo {author} {\bibfnamefont {P.~T.}\ \bibnamefont
  {Dumitrescu}}, \bibinfo {author} {\bibfnamefont {A.}~\bibnamefont
  {Goremykina}}, \bibinfo {author} {\bibfnamefont {S.~A.}\ \bibnamefont
  {Parameswaran}}, \bibinfo {author} {\bibfnamefont {M.}~\bibnamefont
  {Serbyn}},\ and\ \bibinfo {author} {\bibfnamefont {R.}~\bibnamefont
  {Vasseur}},\ }\bibfield  {title} {\bibinfo {title} {Kosterlitz-thouless
  scaling at many-body localization phase transitions},\ }\href
  {https://doi.org/10.1103/PhysRevB.99.094205} {\bibfield  {journal} {\bibinfo
  {journal} {Phys. Rev. B}\ }\textbf {\bibinfo {volume} {99}},\ \bibinfo
  {pages} {094205} (\bibinfo {year} {2019})}\BibitemShut {NoStop}%
\bibitem [{\citenamefont {Goremykina}\ \emph {et~al.}(2019)\citenamefont
  {Goremykina}, \citenamefont {Vasseur},\ and\ \citenamefont
  {Serbyn}}]{Goremykina19}%
  \BibitemOpen
  \bibfield  {author} {\bibinfo {author} {\bibfnamefont {A.}~\bibnamefont
  {Goremykina}}, \bibinfo {author} {\bibfnamefont {R.}~\bibnamefont
  {Vasseur}},\ and\ \bibinfo {author} {\bibfnamefont {M.}~\bibnamefont
  {Serbyn}},\ }\bibfield  {title} {\bibinfo {title} {Analytically solvable
  renormalization group for the many-body localization transition},\ }\href
  {https://doi.org/10.1103/PhysRevLett.122.040601} {\bibfield  {journal}
  {\bibinfo  {journal} {Phys. Rev. Lett.}\ }\textbf {\bibinfo {volume} {122}},\
  \bibinfo {pages} {040601} (\bibinfo {year} {2019})}\BibitemShut {NoStop}%
\bibitem [{\citenamefont {Aramthottil}\ \emph {et~al.}(2021)\citenamefont
  {Aramthottil}, \citenamefont {Chanda}, \citenamefont {Sierant},\ and\
  \citenamefont {Zakrzewski}}]{Aramthottil21}%
  \BibitemOpen
  \bibfield  {author} {\bibinfo {author} {\bibfnamefont {A.~S.}\ \bibnamefont
  {Aramthottil}}, \bibinfo {author} {\bibfnamefont {T.}~\bibnamefont {Chanda}},
  \bibinfo {author} {\bibfnamefont {P.}~\bibnamefont {Sierant}},\ and\ \bibinfo
  {author} {\bibfnamefont {J.}~\bibnamefont {Zakrzewski}},\ }\bibfield  {title}
  {\bibinfo {title} {Finite-size scaling analysis of the many-body localization
  transition in quasiperiodic spin chains},\ }\href
  {https://doi.org/10.1103/PhysRevB.104.214201} {\bibfield  {journal} {\bibinfo
   {journal} {Phys. Rev. B}\ }\textbf {\bibinfo {volume} {104}},\ \bibinfo
  {pages} {214201} (\bibinfo {year} {2021})}\BibitemShut {NoStop}%
\end{thebibliography}%
\end{document}